\documentclass[
reprint,
superscriptaddress,
 amsmath,amssymb,
 aps,
]{revtex4-2}

\usepackage{graphicx}
\usepackage{dcolumn}
\usepackage{bm}
\usepackage{hyperref}

\begin{document}

\preprint{APS/123-QED}

\title{Lattice-plasmon-induced asymmetric transmission\\ in two-dimensional
chiral arrays}

\title{Lattice-plasmon-induced asymmetric transmission in two-dimensional
chiral arrays}
\author{N. Apurv Chaitanya}
\affiliation{Tecnologico de Monterrey, School of Engineering and Sciences, Ave. Eugenio
Garza Sada 2501, Monterrey, N.L., Mexico, 64849.}
\author{M. A. Butt}
\affiliation{Max Planck Institute for the Science of Light, Staudtstrasse 2, D-91058
Erlangen, Germany }
\affiliation{Institute of Optics, Information and Photonics, University Erlangen-Nuremberg,
Staudtstrasse 7/B2, D-91058 Erlangen, Germany}
\affiliation{School in Advanced Optical Technologies, University Erlangen-Nuremberg,
Paul-Gordan-Strasse 6, D-91052 Erlangen, Germany}
\affiliation{Max Planck, University of Ottawa Centre for Extreme and Quantum Photonics,
University of Ottawa, Ottawa, ON, K1N6N5 Canada}
\author{O. Reshef}
\affiliation{Department of Physics, University of Ottawa, Ottawa, Ontario K1N 6N5,
Canada}
\author{Robert W. Boyd}
\affiliation{Department of Physics, University of Ottawa, Ottawa, Ontario K1N 6N5,
Canada}
\affiliation{Max Planck, University of Ottawa Centre for Extreme and Quantum Photonics,
University of Ottawa, Ottawa, ON, K1N6N5 Canada}
\affiliation{The Institute of Optics, University of Rochester, Rochester, NY, 14627,
USA}
\author{P. Banzer}
\affiliation{Max Planck Institute for the Science of Light, Staudtstrasse 2, D-91058
Erlangen, Germany }
\affiliation{Institute of Optics, Information and Photonics, University Erlangen-Nuremberg,
Staudtstrasse 7/B2, D-91058 Erlangen, Germany }
\affiliation{Max Planck, University of Ottawa Centre for Extreme and Quantum Photonics,
University of Ottawa, Ottawa, ON, K1N6N5 Canada}
\affiliation{Institute of Physics, University of Graz, NAWI Graz, Universitätsplatz
5, 8010 Graz, Austria}
\author{Israel De Leon}
\email{ideleon@tec.mx}

\affiliation{Tecnologico de Monterrey, School of Engineering and Sciences, Ave. Eugenio
Garza Sada 2501, Monterrey, N.L., Mexico, 64849.}
\affiliation{Max Planck, University of Ottawa Centre for Extreme and Quantum Photonics,
University of Ottawa, Ottawa, ON, K1N6N5 Canada}
\affiliation{School of Electrical Engineering and Computer Science, University
of Ottawa, Ottawa, ON, K1N6N5 Canada}

\date{\today}

\begin{abstract}
Asymmetric transmission -- direction-selective control of electromagnetic transmission between two ports -- is an important phenomenon typically exhibited by two-dimensional chiral systems. Here, we study this phenomenon in chiral plasmonic metasurfaces supporting lattice plasmons modes. We show, both numerically and experimentally, that asymmetric transmission can be achieved through an unbalanced excitation of such lattice modes by circularly polarized light of opposite handedness. The excitation efficiencies of the lattice modes, and hence the strength of the asymmetric transmission, can be controlled by engineering the in-plane scattering of the individual plasmonic nanoparticles such that the maximum scattering imbalance occurs along one of the in-plane diffraction orders of the metasurface. Our study also shows that, contrary to the case of a non-diffractive metasurface, the lattice-plasmon-enabled asymmetric transmission can occur at normal incidence for cases where the metasurface is composed of chiral or achiral nanoparticles possessing 4-fold rotational symmetry.
\end{abstract}

\maketitle


\section{Introduction}
\vspace{-3mm}

Chirality refers to an intrinsic sense of handedness of a three-dimensional (3D) structure, which remains invariant regardless of the direction of observation. Because of this property, the structure and its mirror image cannot be brought into congruence by a translation and rotation operation \citep{wagniere2008chirality}. Proteins, chemical systems and many bio-molecules are known to be chiral, exhibiting contrasting functions and properties between its mirror-symmetric pair (enantiomer)\citep{Maestre1982,yoon2003privileged,pfaltz2004design}. 3D chiral systems yield a different optical response when interacting with either handedness of circularly polarized light, leading to phenomena such as circular dichrosim and optical activity.  Both phenomena are widely used for characterizing the optical response of chiral systems \citep{berova2000circular}. Chirality could also be defined for two-dimensional (2D) structures. However, as planar structures do not possess an intrinsic handedness, chirality is defined by the inability of a 2D structure and its mirror image to be brought into congruence unless lifted from its plane \citep{Papakostas2003}. Systems composed of achiral structures are known to exhibit chiral effects as well, provided that the surface normal and the electric field vector form a chiral triad \citep{Papakostas2003,Plum2009,DeLeon2015}. This phenomenon, usually called \emph{extrinsic chirality}, is indistinguishable from standard (intrinsic) chirality \citep{Extrinsic_chirality}, and has been reported in a variety of systems, including single nanoparticles \citep{C4NR04433A,PhysRevX.4.011005}.

\begin{figure*}
\includegraphics[width=12cm]{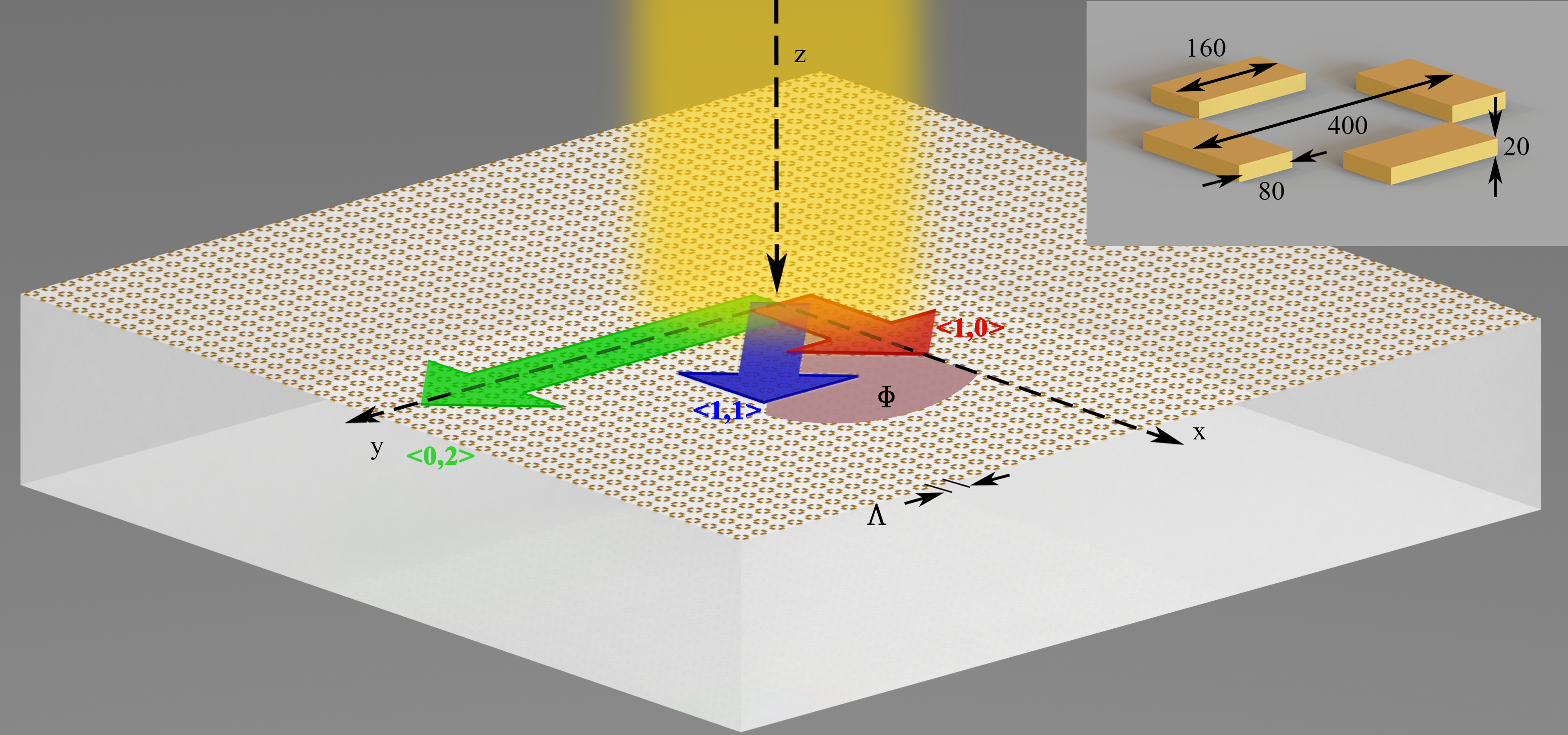}\caption{Schematic illustration of the  metasurfaces under investigation. The metasurface is  illuminated at normal incidence by circularly polarized light propagating in the positive $z$ direction. The metasurface consists of a square array of gold quadrumers embedded in a homogeneous medium. The arrangement and  dimensions of the four nanoantennas composing the quadrumer are shown in the inset. All dimension are given in nanometers. \label{fig:Scheme}}
\end{figure*}

The sense of handedness of a 2D chiral structure is reversed when it is flipped with respect to an in-plane axis \citep{Extrinsic_chirality}. Because of this property, 2D chiral structures can show asymmetric optical transmission, i.e., for a given circular polarization state, the optical transmittance depends on the side of the sample that is being illuminated \citep{Fedotov2006,Fedotov2007,Schwanecke2008,Aba2018}. This is in compliance with the Lorentz reciprocity theorem and does not require magnetic fields as for the Faraday effect \citep{Fedotov2006}. The phenomenon can be also observed as a difference in reflection and/or a difference in absorption. Previous reports of asymmetric transmission in planar arrays of chiral elements lacking 4-fold rotational symmetry have attributed the phenomenon to the simultaneous presence of \emph{anisotropy} and \emph{losses} \citep{Fedotov2006,Fedotov2007,khanikaev2016experimental}. When excited at normal incidence, 4-fold symmetric chiral holes appear not to exhibit any asymmetric transmission \citep{krasavin2005polarization,Krasavin_2006,Reichelt2006}. However, this is not necessarily true for a periodic arrangement of such chiral nanostructures, because of the possibility of loss channels associated with higher-order diffraction \citep{PhysRevE.71.037603}. 

Chiroptical phenomena have been extensively explored in plasmonic systems such as individual nanostructures \citep{Meinzer2013,Hopkins2015,banzer2016chiral,wozniak2018chiroptical}, 2D array of plasmonic nanostructures (metasurfaces) \citep{Drezet:08,PhysRevX.2.031010,Valev2013,zu2016planar} and plasmonic nanostructures in suspension (metafluids) \citep{schreiber2013chiral}. Of particular interest is the transmittance for circular polarized excitation of plasmonic metasurfaces with periodic inter-particle spacing of the order of the optical wavelength. These structures can support lattice plasmon modes  enabled by  diffractive coupling of localized plasmons \citep{Kravets2018,bin2021ultra}. The propagation losses of lattice plasmon modes can act as a channel for \emph{losses}, while the topography of the individual plasmonic nanostructure or the array as a whole can act as a source of \emph{anisotropy}, both fundamental for asymmetric transmission \citep{Fedotov2007}. While previous works on asymmetric transmission in 2D chiral structures have greatly contributed to our understanding of low-dimensional chirality, an investigation of the effects of lattice plasmon modes in metasurfaces is still lacking. 

In this work we show, both numerically and experimentally, that lattice plasmon modes can play a key role in enabling asymmetric transmission in diffractive chiral metasurfaces. This asymmetric transmission mechanism relies on different lattice plasmon mode excitation efficiencies for left circularly polarized (LCP) and right circularly polarized (RCP) light. Such excitation efficiencies can be controlled by tailoring the nanoparticle’s in-plane distribution of scattered light for circularly polarized excitation and its alignment with the in-plane diffraction orders of the metasurface. Contrary to the case of non-diffractive chiral metasurfaces, the lattice-plasmon assisted mechanism described here leads to asymmetric transmission even for nanostructures with 4-fold rotational symmetry and normal incidence illumination. For simplicity we analyze this phenomenon for  a metasurface  composed of  an array of achiral nanoparticles - a \textit{quadrumer} consisting of four nanoantennas as shown in inset of Fig. \ref{fig:Scheme}. We show that near the Rayleigh anomaly condition, where the asymmetric transmission effect is strongest, the metasurface supports lattice plasmon modes that respond selectively to the polarization handedness, confirming the underlying asymmetric transmission mechanism enabled by the lattice plasmon modes. These results are of general relevance for understanding the role of diffracted waves in 2D chiral systems and in particular for the design of chiral plasmonic metasurfaces. 

\begin{figure*}
\includegraphics[width=15cm]{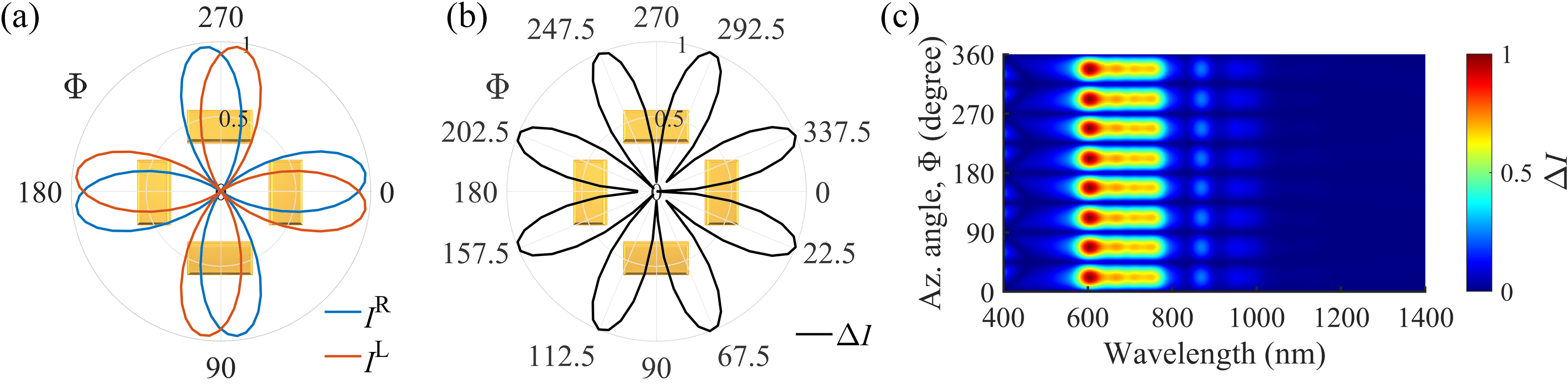}\caption{Optical response of isolated nanostructure obtained through FDTD simulations. (a) The polar plot of the far field scattered intensity as a function of in-plane angle for an individual nanostructure. $I^{{\rm R}}$, $I^{{\rm L}}$ are the normalized scattered intensity for RCP (solid blue) and LCP (solid red) excitation. (b) The normalized plots of differential in-plane scattering, $\Delta I=\left|I^{{\rm R}}-I^{{\rm L}}\right|,$ for 620 nm excitation wavelength. Superimposed on  the plots are the isolated nanostructure for  reference. (c) The variation of $\Delta I$ as a function of wavelength for the nanostructure, in normalized units.\label{fig:CDS} }
\end{figure*}

\vspace{-5mm}
\section{Asymmetric transmission mechanism and numerical analysis}
\vspace{-3mm}

Consider the metasurface depicted in Fig.~\ref{fig:Scheme}, which consists of a planar array of quadrumer nanostructures. The quadrumers are made of gold, arranged into a squared lattice, and surrounded by a homogeneous medium with refractive index $n=1.51$ (i.e., glass). The array lies on the ($x$, $y$) plane, having a lattice spacing of $\Lambda=600$ nm. The dimensions of each quadrumer are indicated in the inset of Fig.~\ref{fig:Scheme}. The metasurface is illuminated by a circularly polarized planewave propagating in the positive $z$ direction and impinging at normal incidence onto the surface. Note that  although the individual quadrumers are achiral, it is possible to create a 2D-chiral metasurface by rotating the quadrumer at each lattice point around its center.

The metasurface can support lattice plasmon modes when $\Lambda$ is of the order of the optical wavelength. These modes result from the coupling of light scattered from neighboring particles through grazing diffraction orders, or Rayleigh anomalies \citep{Kravets2008,Kravets2018}. For normally incident excitation, such Rayleigh anomalies occur at wavelengths given by \citep{khlopin2017lattice}, 
\begin{equation}
\lambda_{p,q}=n\Lambda\frac{\sqrt{\left(p^{2}+q^{2}\right)-q^{2}}\pm p}{p^{2}+q^{2}},\label{eq:RA}
\end{equation}
where $p$ and $q$ are integers indicating the in-plane diffraction orders along the $x$ and $y$ directions, respectively. The Rayleigh anomaly condition for the first diffraction order, $\left\langle p,q\right\rangle =\left\langle 0,1\right\rangle =\left\langle 1,0\right\rangle $, is degenerate and occurs at the wavelength $\lambda_{1,0}=906$ nm. Similarly, the next Rayleigh anomaly, which corresponds to the diffraction order $\left\langle 1,1\right\rangle$, occurs at $\lambda_{1,1}=640$ nm. Lattice plasmon modes propagating along the different in-plane $\left\langle p,q\right\rangle $ diffraction orders are supported at wavelengths close to those described by Eq.~(1). The direction of the three main $\left\langle p,q\right\rangle $ diffraction orders are depicted in Fig.~\ref{fig:Scheme}. 

To understand the asymmetric transmission mechanism exhibited by this metasurface, we first analyze the optical scattering properties of the individual quadrumers for LCP and RCP plane-wave illumination. For this, we use a fully vectorial Maxwell equation solver based on the finite difference time domain (FDTD) method. The permittivity of gold was taken from Ref.~\citep{Johnson1972}. Fig.~\ref{fig:CDS}a shows the obtained in-plane angular distribution of the scattered intensity at the wavelength $\lambda=620$ nm, which is close to the particle's resonance. Here, $I^{{\rm L}}$ and $I^{{\rm R}}$ are the in-plane scattered intensities for LCP and RCP excitation, respectively. The difference between these intensities, $\Delta I=|I^{{\rm R}}-I^{{\rm L}}|$, will be referred to as the differential in-plane scattering. This quantity is plotted in Fig.~\ref{fig:CDS}b for the same wavelength as Fig.~\ref{fig:CDS}a, and as a function of   wavelength in Fig.~\ref{fig:CDS}c. Observe that regardless of the optical wavelength, the differential in-plane scattering has maxima ($\Delta I_{\rm max}$) and minima ($\Delta I_{\rm min}$) at particular azimuthal angles, whose values are odd and even integer multiples of $22.5^\circ$, respectively.

\begin{figure*}
\includegraphics[width=16cm]{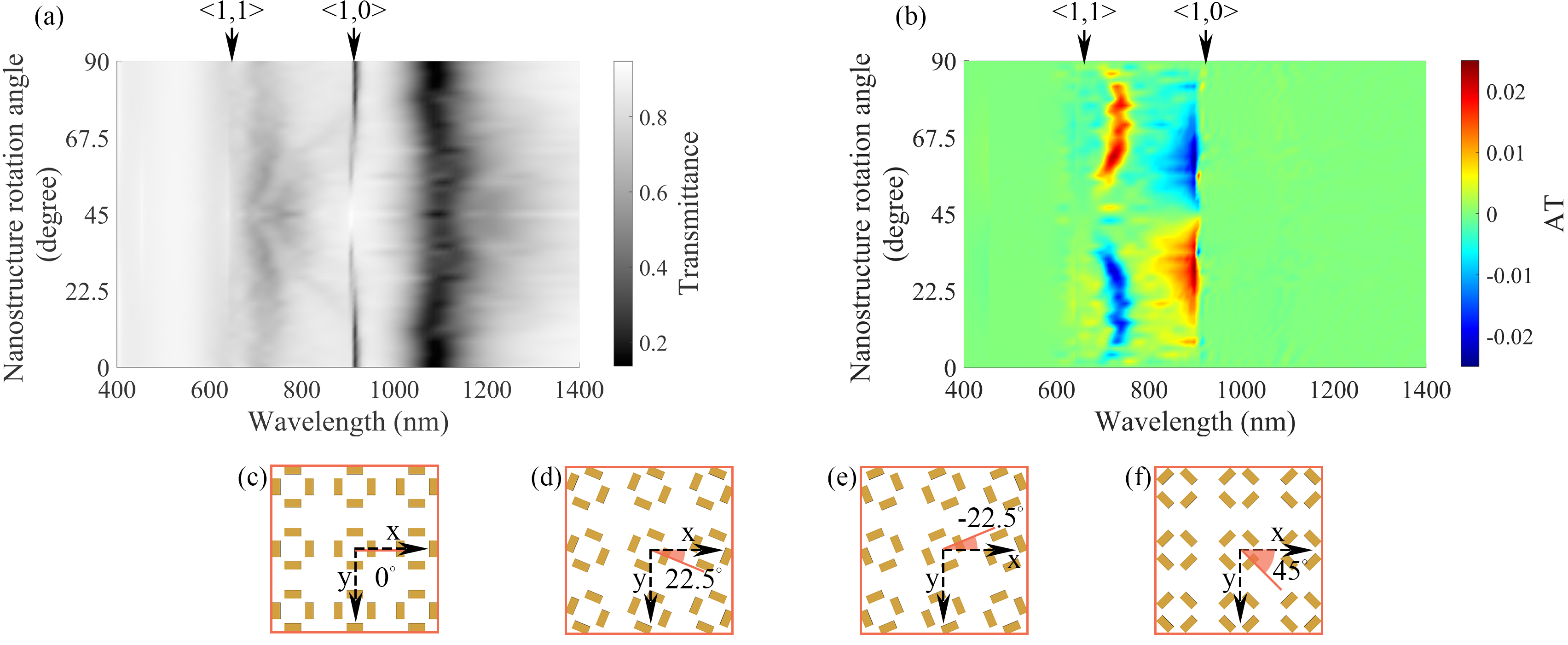}\caption{(a) Numerically calculated transmittance spectra for RCP excitation for the metasurface as a function of rotation of the individual nanostructure. The arrows on the top of the results denote the wavelengths at which the Rayleigh anomalies occur, corresponding to a lattice period of $\Lambda=600$ nm in a homogeneous medium with $n=1.51$.  (b) The variation in asymmetric transmission calculated using Eq. \ref{eq:AT} as a function of the quadrumer's rotation angle. Asymmetric transmission occurs when the maxima of the differential in-plane scattering are aligned along the propagation direction of the grazing diffraction order. (c-f) Show the schematic illustration of the metasurface when the individual nanostructure  at each lattice point is rotated around its center. \label{fig:AT Sim}}
\end{figure*}

The individual quadrumer exhibits a non-zero differential in-plane scattering in a spectral range overlapping with the Rayleigh anomalies of the array. Consequently, the lattice plasmon excitation efficiency will vary depending on whether the quadrumer array is illuminated with RCP or LCP light, leading to different transmittances for each case. In turn, this leads to asymmetric transmission, as reversing the polarization handedness is equivalent to illuminating the metasurface from the opposite side. Moreover, since $\Delta I$ varies along the in-plane angle, the differential excitation of lattice plasmon modes by RCP and LCP light also vary with the relative angle between $\Delta I$ and the lattice plasmon mode propagation direction. In particular, when $\Delta I_{{\rm max}}$ ($\Delta I_{{\rm min}}$) is along the propagation direction of a lattice plasmon mode, then the difference in efficiency of excitation of that lattice plasmon mode will be maximized (minimized).

To analyze this effect, the transmittance of the metasurface as a function of quadrumer's rotation angles is computed and plotted in Fig.~\ref{fig:AT Sim}a for RCP illumination; a similar result is obtained for LCP illumination (not shown). The relative orientation of the nanostructure for different rotation angles with respect to the lattice is illustrated in Figs.~\ref{fig:AT Sim}c-f. Fig.~\ref{fig:AT Sim}b shows the asymmetric transmission calculated from the numerical results as
\begin{equation}
AT=\frac{T^{{\rm R}}-T^{{\rm L}}}{T^{{\rm R}}+T^{{\rm L}}},\label{eq:AT}
\end{equation}
where $T^{{\rm R}}$ and $T^{{\rm L}}$ are the transmittance for RCP and LCP excitation, respectively. The asymmetric transmission vanishes for wavelengths beyond 906 nm because the Rayleigh anomaly condition is not satisfied, and hence lattice plasmon modes are not supported. On the other hand,  the asymmetric transmission is maximized at wavelengths close to the Rayleigh anomaly conditions ($\lambda_{10}$ and $\lambda_{11}$), where lattice plasmon modes are most efficiently excited. 

Most importantly, we note that the asymmetric transmission depends on the in-plane rotation of the individual quadrumers. Clearly, the largest asymmetric transmission is obtained when the individual nanostructure is rotated by odd integer multiples of $22.5^{\circ}$, which coincides with the angles for $\Delta I_{{\rm max}}$ of the isolated nanostructure (see Fig. \ref{fig:CDS}b). Detailed analysis of the asymmetric transmission arising from an array of chiral nanostructures - \textit{gammadia} is given in the supplementary information. From these results, we conclude that the value of the differential in-plane scattering of individual nanostructure, $\Delta I$ along the diffraction orders at the Rayleigh anomaly condition determines the strength of the asymmetric transmission. Furthermore, we note that when lattice plasmon modes are excited, the metasurface can exhibit asymmetric transmission as long as the individual nanostructure exhibit differential in-plane scattering along the lattice plasmon mode propagation direction.  As such it is irrelevant whether the nanostructure is chiral or achiral and if it has 4-fold rotational symmetry or not.

Finally, an analysis of the transmission characteristics for RCP and LCP excitation along different diffraction orders, away from the Rayleigh anomaly condition, reveals that the transmission into  the zeroth order is degenerate for both excitations irrespective of the wavelength and nanostructure rotation angles. Thus, the asymmetric transmission observed in Fig. \ref{fig:AT Sim}b, is caused solely by the difference in transmission of higher diffraction orders. 

\begin{figure*}
\includegraphics[width=19cm]{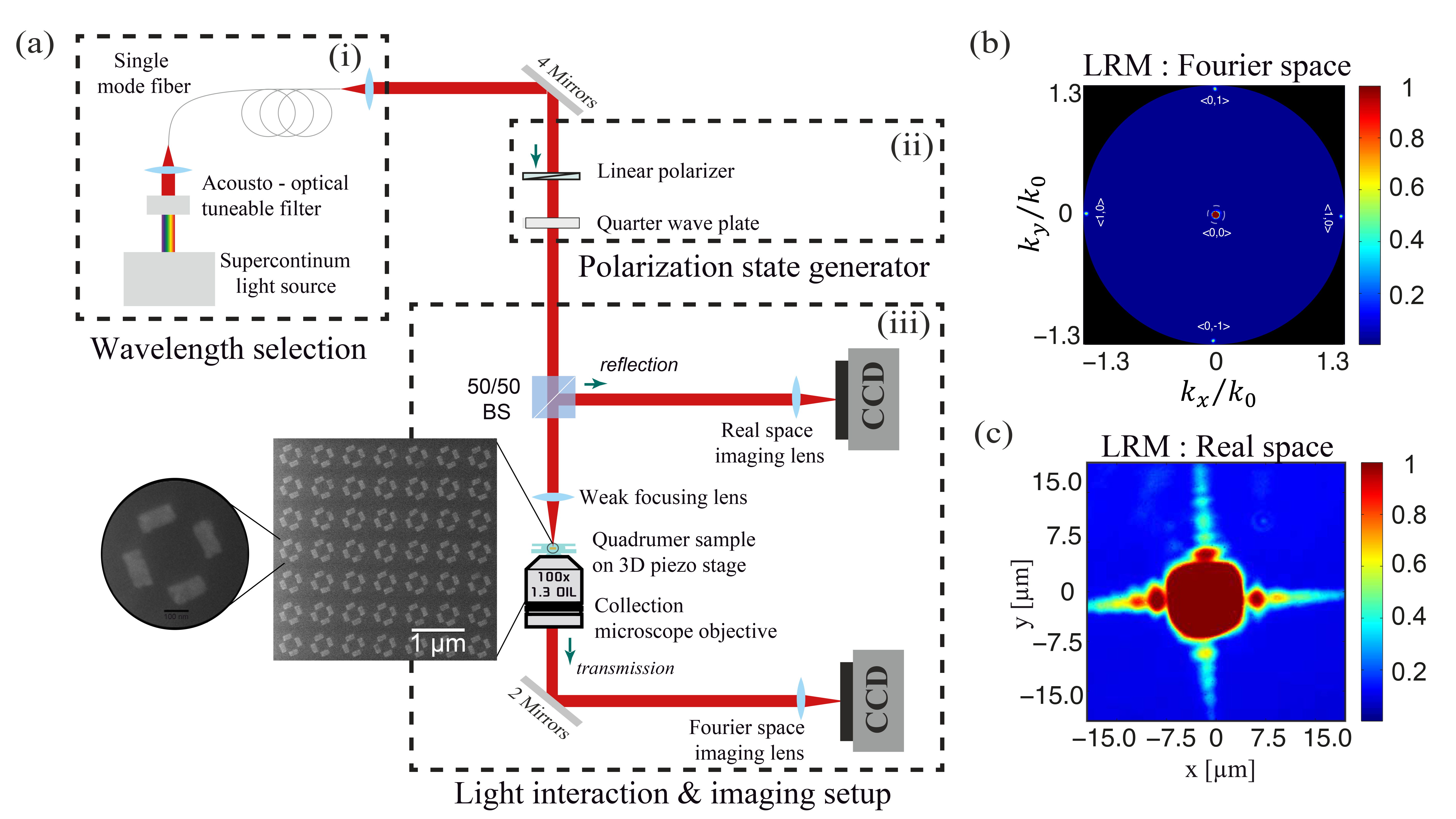}\caption{(a) Sketch of the experimental setup. A broadband light source is  filtered to perform spectrally resolved measurements. A broadband polarizer followed by a quarter-wave plate are used to prepare the polarization state of the excitation beam. The beam is weakly focused, producing a focal spot of approximately 50 $\mu$m diameter on the quadrumer metasurface. The transmitted light is collected with a microscope objective of $\mathrm{NA}=1.3$. The objective's back focal plane, containing the $k$-space information, is imaged onto a CCD camera. (b) The $k$-space image recorded in transmission. Zeroth and first diffraction orders are visible, which were evaluated for contributions to asymmetric transmission. (c) Direct imaging of the quadrumer metasurface in reflection at a wavelength near the Rayleigh anomaly, clearly showing the leaky propagating first diffraction orders. \label{fig:EXP1}} 
\end{figure*}

\vspace{-5mm}
\section{Experimental results}
\vspace{-3mm}

Based on the numerical analysis, the fabrication of quadrumer metasurfaces with required orientation of individual structures was carried out. Details about the  fabrication process can be found in supplementary information. A schematic illustration of the experimental setup used for the investigation is shown in Fig. \ref{fig:EXP1}a. A filtered broadband light source is used to carry out spectrally resolved measurements in a range of 630 nm~--~950 nm. A broadband polarizer followed by a quarter-wave plate are used to prepare the polarization state of the excitation beam. The beam is then weakly focused to a diameter of approximately 50 $\mu$m with the help of a convex lens ($f=300$ mm), and the metasurface is precisely placed in the focal spot with the help of a 3-axis piezo stage. Detailed description of experimental setup can be found in supplementary information.

We consider three metasurfaces with in-plane rotation of nanostructures at $22.5^{\circ}$, $-22.5^{\circ}$ and $45^{\circ}$ to measure the lattice-plasmon induced asymmetric transmission. For the present metasurface designs, a significant fraction of the transmitted power remains in the zeroth order, hindering the experimental observations of the effect. Hence, we perform diffraction-resolved measurements to identify the independent power contributions of the transmitted zeroth and first diffraction orders. To this end, the transmitted light is collected with a microscope objective ($\mathrm{NA}=1.3$), with its back focal plane imaged onto a CCD camera. The back focal plane gives us access to the wave-vector space ($k$-space) of the transmitted light from which we obtain the diffraction-resolved power contributions  (see supplementary information). The zeroth order is measured for all wavelengths. The first diffraction orders can be measured only for wavelengths shorter than 780~nm because of the limitation imposed by the maximum collection angle of the microscope objective ($60.25^{\circ}$). This limitation prevents us from measuring the power coupled to lattice plasmon modes (near the Rayleigh anomaly wavelengths). For such cases, we employ direct imaging of the metasurface in reflection via leakage radiation microscopy. This technique allows us to obtain qualitative information of the power coupled to lattice plasmon modes by visualizing light leaked as they propagate. Figs. \ref{fig:EXP1}b and \ref{fig:EXP1}c illustrate, respectively, typical $k$-space and leakage radiation microscopy images recorded using the above mentioned techniques.

\begin{figure*}
\includegraphics[width=14cm]{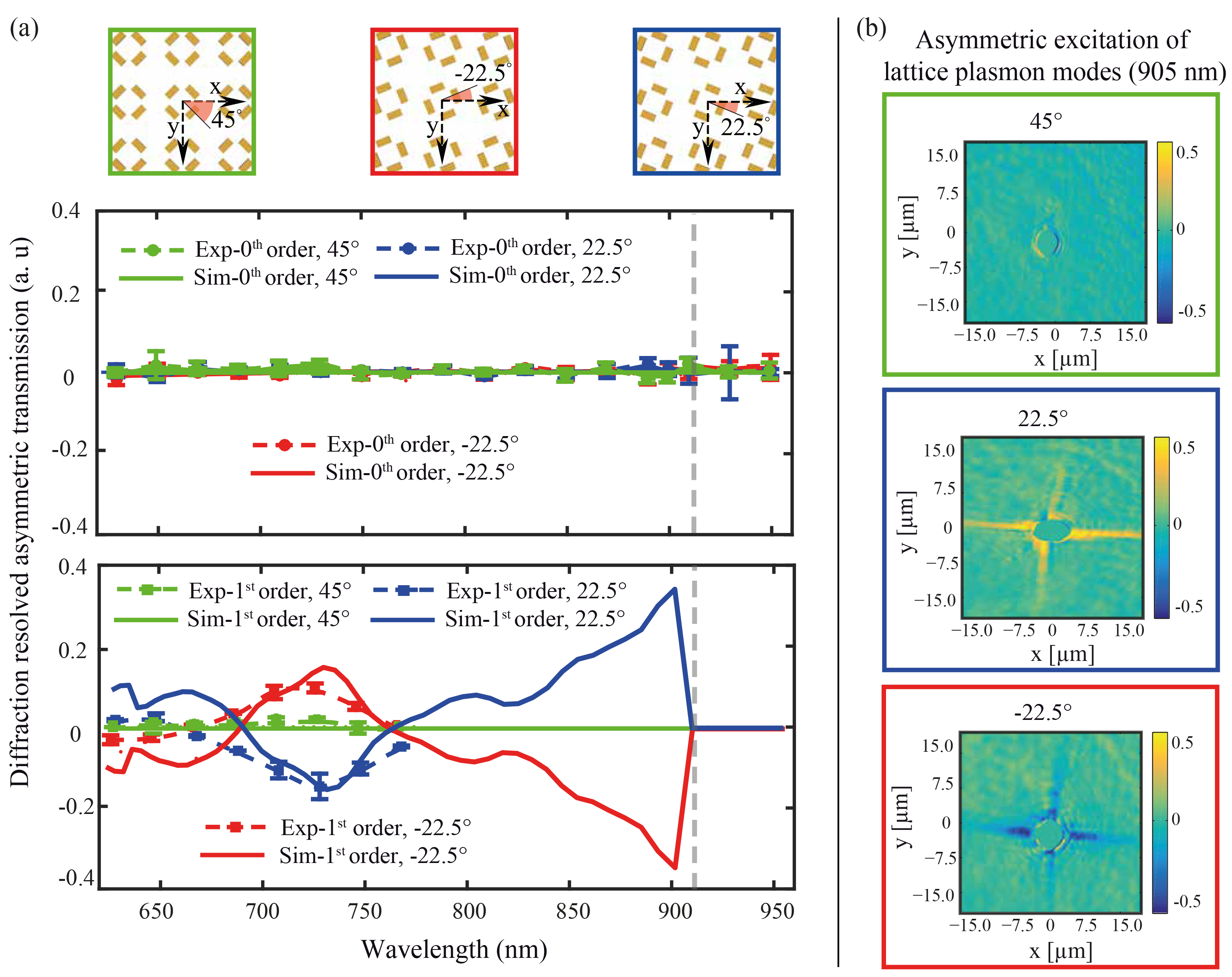}\caption{Experimental and simulated diffraction resolved asymmetric transmission results for achiral ($45^{\circ}$ rotation) and chiral ($22.5^{\circ}$  $-22.5^{\circ}$ rotations) metasurfaces. (a) Wavelength versus  diffraction resolved asymmetric transmission for zeroth  (top panel) and first diffraction order (bottom panel) for the achiral (green) and chiral (blue and red) metasurfaces. (b) Differential leakage radiation microscopy images recorder at a wavelength of 905~nm, showing asymmetric excitation of lattice plasmon modes for chiral ($22.5^{\circ}$  $-22.5^{\circ}$ rotations) metasurfaces with opposing signs, and symmetric excitation of lattice plasmon modes for the achiral metasurface ($45^{\circ}$).
\label{fig:EXP2}} 
\end{figure*}

The experimental measurements of lattice-plasmon induced asymmetric transmission obtained for the achiral ($45^{\circ}$ rotation) and chiral ($22.5^{\circ}$ and $-22.5^{\circ}$ rotations) metasurfaces are summarized in Fig.~\ref{fig:EXP2}. The diffraction resolved asymmetric transmission for the zeroth and first diffraction orders are plotted in Fig.~\ref{fig:EXP2}a as a function of the wavelength; both, measurements (dashed lines) and simulations (solid lines) are included in the plots. Diffraction resolved asymetric transmission is calculated using Eq. \ref{eq:AT}, but  $T_R$ and $T_L$ replaced with $T_R^{\left\langle p,q\right\rangle}$  and $T_L^{\left\langle p,q\right\rangle}$ which are the  transmittance in the $\left\langle p,q\right\rangle$ diffraction order for RCP and LCP excitation, respectively. Leakage radiation microscopy images are used to study the phenomenon at a wavelength close to the Rayleigh anomaly (marked by dashed grey line in spectral plots), by constructing differential measurements obtained from a pixel-by-pixel subtraction of two camera shots taken for excitation with opposite polarization handedness. Such differential leakage radiation microscopy images, shown in Fig.~\ref{fig:EXP2}b, allow us to visualize the asymmetric excitation of lattice plasmon modes responsible for the asymmetric transmission.

First, we discuss the case of the metasurface with quadrumers rotated by $45^{\circ}$ in-plane (see Fig.~\ref{fig:EXP2}a top-left), which constitutes an achiral metasurface. As expected, this metasurface results in vanishing asymmetric transmission in first diffraction orders, which is confirmed by both experimental and simulation results. The differential-leakage radiation microscopy image recorded for this metasurface, also confirms vanishing  asymmetric propagating modes (see Fig.~\ref{fig:EXP2}b, marked in green box). Next, consider the metasurface constituted by quadrumers with in-plane rotations of $22.5^{\circ}$ and $-22.5^{\circ}$ (see Fig.~\ref{fig:EXP2}a top-center and top-right, respectively). As expected from simulation results shown in Fig.~\ref{fig:AT Sim}d, the two corresponding cases show opposing signs of asymmetric transmission for first order. This is confirmed by the experimental results as shown in the spectral plot in Fig. \ref{fig:EXP2}a. The differential leakage radiation microscopy image recorded for the chiral arrangements depict asymmetrically-excited lattice plasmon modes with opposing signs (see Fig.~\ref{fig:EXP2}b, marked in blue and red box).

To demonstrate and validate the sole contribution of the first diffraction orders to the observed asymmetric transmission, we also show the corresponding simulation and experimental results for the zeroth diffraction order not featuring any significant spectral dependence for all three studied metasurfaces (see Fig. \ref{fig:EXP2}a center). The absence of  asymmetric transmission for all three quadrumer metasurfaces for the zeroth order underscores the essential role of lattice plasmon modes in the assymmetric transmission mechanism.

\vspace{-5mm}
\section{Conclusions}
\vspace{-3mm}

In conclusion, we studied the role of lattice plasmon modes on the phenomenon of asymmetric transmission in chiral 2D arrays of plasmonic nanoparticles. We showed that the difference in the nanoparticle’s in-plane scattering of LCP and RCP light can contribute to asymmetric transmission in 2D metasurfaces through the excitation of polarization-handedness selective lattice plasmon modes. The different excitation efficiencies of the lattice plasmon modes, produced by unbalanced in-plane scattering of LCP and RCP light and its coupling to grazing diffraction orders of the periodic lattice, result in asymmetries in transmission, reflection and absorption. As the difference in the in-plane scattered intensity varies as a function of the in-plane angle around the nanostructure, rotating the nanostructure in the plane of the metasurface can act as a handle to control the metasurface’s asymmetric transmission. Using diffraction-order resolved transmission measurements, we demonstrated that the asymmetric transmission results solely from contributions of higher-order diffracted waves, while the zeroth-order transmission in not asymmetric. Furthermore, we showed that, even though the isolated nanostructure has a 4-fold rotational symmetry, the diffractive metasurfaces studied here exhibit asymmetric transmission for normal incidence illumination within the spectral range for which lattice plasmon modes are supported. This behavior is contrary to what occurs in a non-diffractive chiral metasurface. Our study sheds new light on asymmetric or chiral phenomena in metasurfaces and the important role of diffraction effects such as the excitation of lattice plasmon modes.

\begin{acknowledgments}
\vspace{-3mm}

IDL acknowledges the financial support from the Federico Baur Endowed Chair in Nanotechnology. RWB acknowledges support through the Natural Sciences and Engineering Research Council of Canada, the Canada Research Chairs program, the US ARO Grant W911NF-18-1-0337, and US  DARPA Award W911NF1810369
\end{acknowledgments}


\begin{thebibliography}{40}%
	\makeatletter
	\providecommand \@ifxundefined [1]{%
		\@ifx{#1\undefined}
	}%
	\providecommand \@ifnum [1]{%
		\ifnum #1\expandafter \@firstoftwo
		\else \expandafter \@secondoftwo
		\fi
	}%
	\providecommand \@ifx [1]{%
		\ifx #1\expandafter \@firstoftwo
		\else \expandafter \@secondoftwo
		\fi
	}%
	\providecommand \natexlab [1]{#1}%
	\providecommand \enquote  [1]{``#1''}%
	\providecommand \bibnamefont  [1]{#1}%
	\providecommand \bibfnamefont [1]{#1}%
	\providecommand \citenamefont [1]{#1}%
	\providecommand \href@noop [0]{\@secondoftwo}%
	\providecommand \href [0]{\begingroup \@sanitize@url \@href}%
	\providecommand \@href[1]{\@@startlink{#1}\@@href}%
	\providecommand \@@href[1]{\endgroup#1\@@endlink}%
	\providecommand \@sanitize@url [0]{\catcode `\\12\catcode `\$12\catcode
		`\&12\catcode `\#12\catcode `\^12\catcode `\_12\catcode `\%12\relax}%
	\providecommand \@@startlink[1]{}%
	\providecommand \@@endlink[0]{}%
	\providecommand \url  [0]{\begingroup\@sanitize@url \@url }%
	\providecommand \@url [1]{\endgroup\@href {#1}{\urlprefix }}%
	\providecommand \urlprefix  [0]{URL }%
	\providecommand \Eprint [0]{\href }%
	\providecommand \doibase [0]{https://doi.org/}%
	\providecommand \selectlanguage [0]{\@gobble}%
	\providecommand \bibinfo  [0]{\@secondoftwo}%
	\providecommand \bibfield  [0]{\@secondoftwo}%
	\providecommand \translation [1]{[#1]}%
	\providecommand \BibitemOpen [0]{}%
	\providecommand \bibitemStop [0]{}%
	\providecommand \bibitemNoStop [0]{.\EOS\space}%
	\providecommand \EOS [0]{\spacefactor3000\relax}%
	\providecommand \BibitemShut  [1]{\csname bibitem#1\endcsname}%
	\let\auto@bib@innerbib\@empty
	\bibitem [{\citenamefont {Wagni{\`e}re}(2008)}]{wagniere2008chirality}%
	\BibitemOpen
	\bibfield  {author} {\bibinfo {author} {\bibfnamefont {G.~H.}\ \bibnamefont
			{Wagni{\`e}re}},\ }\href@noop {} {\emph {\bibinfo {title} {On chirality and
				the universal asymmetry: reflections on image and mirror image}}}\ (\bibinfo
	{publisher} {John Wiley \& Sons},\ \bibinfo {year} {2008})\BibitemShut
	{NoStop}%
	\bibitem [{\citenamefont {Maestre}\ \emph {et~al.}(1982)\citenamefont
		{Maestre}, \citenamefont {Bustamante}, \citenamefont {Hayes}, \citenamefont
		{Subirana},\ and\ \citenamefont {Tinoco}}]{Maestre1982}%
	\BibitemOpen
	\bibfield  {author} {\bibinfo {author} {\bibfnamefont {M.~F.}\ \bibnamefont
			{Maestre}}, \bibinfo {author} {\bibfnamefont {C.}~\bibnamefont {Bustamante}},
		\bibinfo {author} {\bibfnamefont {T.~L.}\ \bibnamefont {Hayes}}, \bibinfo
		{author} {\bibfnamefont {J.~A.}\ \bibnamefont {Subirana}},\ and\ \bibinfo
		{author} {\bibfnamefont {I.}~\bibnamefont {Tinoco}},\ }\href
	{https://doi.org/10.1038/298773a0} {\bibinfo {title} {{Differential
				scattering of circularly polarized light by the helical sperm head from the
				octopus Eledone cirrhosa}}} (\bibinfo {year} {1982})\BibitemShut {NoStop}%
	\bibitem [{\citenamefont {Yoon}\ and\ \citenamefont
		{Jacobsen}(2003)}]{yoon2003privileged}%
	\BibitemOpen
	\bibfield  {author} {\bibinfo {author} {\bibfnamefont {T.~P.}\ \bibnamefont
			{Yoon}}\ and\ \bibinfo {author} {\bibfnamefont {E.~N.}\ \bibnamefont
			{Jacobsen}},\ }\bibfield  {title} {\bibinfo {title} {Privileged chiral
			catalysts},\ }\href@noop {} {\bibfield  {journal} {\bibinfo  {journal}
			{Science}\ }\textbf {\bibinfo {volume} {299}},\ \bibinfo {pages} {1691}
		(\bibinfo {year} {2003})}\BibitemShut {NoStop}%
	\bibitem [{\citenamefont {Pfaltz}\ and\ \citenamefont
		{Drury}(2004)}]{pfaltz2004design}%
	\BibitemOpen
	\bibfield  {author} {\bibinfo {author} {\bibfnamefont {A.}~\bibnamefont
			{Pfaltz}}\ and\ \bibinfo {author} {\bibfnamefont {W.~J.}\ \bibnamefont
			{Drury}},\ }\bibfield  {title} {\bibinfo {title} {Design of chiral ligands
			for asymmetric catalysis: From c2-symmetric p, p-and n, n-ligands to
			sterically and electronically nonsymmetrical p, n-ligands},\ }\href@noop {}
	{\bibfield  {journal} {\bibinfo  {journal} {Proceedings of the National
				Academy of Sciences}\ }\textbf {\bibinfo {volume} {101}},\ \bibinfo {pages}
		{5723} (\bibinfo {year} {2004})}\BibitemShut {NoStop}%
	\bibitem [{\citenamefont {Berova}\ \emph {et~al.}(2000)\citenamefont {Berova},
		\citenamefont {Nakanishi}, \citenamefont {Woody},\ and\ \citenamefont
		{Woody}}]{berova2000circular}%
	\BibitemOpen
	\bibfield  {author} {\bibinfo {author} {\bibfnamefont {N.}~\bibnamefont
			{Berova}}, \bibinfo {author} {\bibfnamefont {K.}~\bibnamefont {Nakanishi}},
		\bibinfo {author} {\bibfnamefont {R.~W.}\ \bibnamefont {Woody}},\ and\
		\bibinfo {author} {\bibfnamefont {R.}~\bibnamefont {Woody}},\ }\href@noop {}
	{\emph {\bibinfo {title} {Circular dichroism: principles and applications}}}\
	(\bibinfo  {publisher} {John Wiley \& Sons},\ \bibinfo {year}
	{2000})\BibitemShut {NoStop}%
	\bibitem [{\citenamefont {Papakostas}\ \emph {et~al.}(2003)\citenamefont
		{Papakostas}, \citenamefont {Potts}, \citenamefont {Bagnall}, \citenamefont
		{Prosvirnin}, \citenamefont {Coles},\ and\ \citenamefont
		{Zheludev}}]{Papakostas2003}%
	\BibitemOpen
	\bibfield  {author} {\bibinfo {author} {\bibfnamefont {A.}~\bibnamefont
			{Papakostas}}, \bibinfo {author} {\bibfnamefont {A.}~\bibnamefont {Potts}},
		\bibinfo {author} {\bibfnamefont {D.~M.}\ \bibnamefont {Bagnall}}, \bibinfo
		{author} {\bibfnamefont {S.~L.}\ \bibnamefont {Prosvirnin}}, \bibinfo
		{author} {\bibfnamefont {H.~J.}\ \bibnamefont {Coles}},\ and\ \bibinfo
		{author} {\bibfnamefont {N.~I.}\ \bibnamefont {Zheludev}},\ }\bibfield
	{title} {\bibinfo {title} {{Optical Manifestations of Planar Chirality}},\
	}\href {https://doi.org/10.1103/PhysRevLett.90.107404} {\bibfield  {journal}
		{\bibinfo  {journal} {Physical Review Letters}\ }\textbf {\bibinfo {volume}
			{90}},\ \bibinfo {pages} {4} (\bibinfo {year} {2003})}\BibitemShut {NoStop}%
	\bibitem [{\citenamefont {Plum}\ \emph
		{et~al.}(2009{\natexlab{a}})\citenamefont {Plum}, \citenamefont {Liu},
		\citenamefont {Fedotov}, \citenamefont {Chen}, \citenamefont {Tsai},\ and\
		\citenamefont {Zheludev}}]{Plum2009}%
	\BibitemOpen
	\bibfield  {author} {\bibinfo {author} {\bibfnamefont {E.}~\bibnamefont
			{Plum}}, \bibinfo {author} {\bibfnamefont {X.~X.}\ \bibnamefont {Liu}},
		\bibinfo {author} {\bibfnamefont {V.~A.}\ \bibnamefont {Fedotov}}, \bibinfo
		{author} {\bibfnamefont {Y.}~\bibnamefont {Chen}}, \bibinfo {author}
		{\bibfnamefont {D.~P.}\ \bibnamefont {Tsai}},\ and\ \bibinfo {author}
		{\bibfnamefont {N.~I.}\ \bibnamefont {Zheludev}},\ }\bibfield  {title}
	{\bibinfo {title} {{Metamaterials: Optical activity without chirality}},\
	}\href {https://doi.org/10.1103/PhysRevLett.102.113902} {\bibfield  {journal}
		{\bibinfo  {journal} {Physical Review Letters}\ }\textbf {\bibinfo {volume}
			{102}},\ \bibinfo {pages} {1} (\bibinfo {year}
		{2009}{\natexlab{a}})}\BibitemShut {NoStop}%
	\bibitem [{\citenamefont {{De Leon}}\ \emph {et~al.}(2015)\citenamefont {{De
				Leon}}, \citenamefont {Horton}, \citenamefont {Schulz}, \citenamefont
		{Upham}, \citenamefont {Banzer},\ and\ \citenamefont {Boyd}}]{DeLeon2015}%
	\BibitemOpen
	\bibfield  {author} {\bibinfo {author} {\bibfnamefont {I.}~\bibnamefont {{De
					Leon}}}, \bibinfo {author} {\bibfnamefont {M.~J.}\ \bibnamefont {Horton}},
		\bibinfo {author} {\bibfnamefont {S.~A.}\ \bibnamefont {Schulz}}, \bibinfo
		{author} {\bibfnamefont {J.}~\bibnamefont {Upham}}, \bibinfo {author}
		{\bibfnamefont {P.}~\bibnamefont {Banzer}},\ and\ \bibinfo {author}
		{\bibfnamefont {R.~W.}\ \bibnamefont {Boyd}},\ }\bibfield  {title} {\bibinfo
		{title} {{Strong, spectrally-tunable chirality in diffractive
				metasurfaces}},\ }\bibfield  {journal} {\bibinfo  {journal} {Scientific
			Reports}\ }\textbf {\bibinfo {volume} {5}},\ \href
	{https://doi.org/10.1038/srep13034} {10.1038/srep13034} (\bibinfo {year}
	{2015})\BibitemShut {NoStop}%
	\bibitem [{\citenamefont {Plum}\ \emph
		{et~al.}(2009{\natexlab{b}})\citenamefont {Plum}, \citenamefont {Fedotov},\
		and\ \citenamefont {Zheludev}}]{Extrinsic_chirality}%
	\BibitemOpen
	\bibfield  {author} {\bibinfo {author} {\bibfnamefont {E.}~\bibnamefont
			{Plum}}, \bibinfo {author} {\bibfnamefont {V.~A.}\ \bibnamefont {Fedotov}},\
		and\ \bibinfo {author} {\bibfnamefont {N.~I.}\ \bibnamefont {Zheludev}},\
	}\bibfield  {title} {\bibinfo {title} {Extrinsic electromagnetic chirality in
			metamaterials},\ }\href {http://stacks.iop.org/1464-4258/11/i=7/a=074009}
	{\bibfield  {journal} {\bibinfo  {journal} {Journal of Optics A: Pure and
				Applied Optics}\ }\textbf {\bibinfo {volume} {11}},\ \bibinfo {pages}
		{074009} (\bibinfo {year} {2009}{\natexlab{b}})}\BibitemShut {NoStop}%
	\bibitem [{\citenamefont {Lu}\ \emph {et~al.}(2014)\citenamefont {Lu},
		\citenamefont {Wu}, \citenamefont {Zhu}, \citenamefont {Zhao}, \citenamefont
		{Wang}, \citenamefont {Zhan},\ and\ \citenamefont {Ni}}]{C4NR04433A}%
	\BibitemOpen
	\bibfield  {author} {\bibinfo {author} {\bibfnamefont {X.}~\bibnamefont
			{Lu}}, \bibinfo {author} {\bibfnamefont {J.}~\bibnamefont {Wu}}, \bibinfo
		{author} {\bibfnamefont {Q.}~\bibnamefont {Zhu}}, \bibinfo {author}
		{\bibfnamefont {J.}~\bibnamefont {Zhao}}, \bibinfo {author} {\bibfnamefont
			{Q.}~\bibnamefont {Wang}}, \bibinfo {author} {\bibfnamefont {L.}~\bibnamefont
			{Zhan}},\ and\ \bibinfo {author} {\bibfnamefont {W.}~\bibnamefont {Ni}},\
	}\bibfield  {title} {\bibinfo {title} {Circular dichroism from single
			plasmonic nanostructures with extrinsic chirality},\ }\href
	{https://doi.org/10.1039/C4NR04433A} {\bibfield  {journal} {\bibinfo
			{journal} {Nanoscale}\ }\textbf {\bibinfo {volume} {6}},\ \bibinfo {pages}
		{14244} (\bibinfo {year} {2014})}\BibitemShut {NoStop}%
	\bibitem [{\citenamefont {Yokoyama}\ \emph {et~al.}(2014)\citenamefont
		{Yokoyama}, \citenamefont {Yoshida}, \citenamefont {Ishii},\ and\
		\citenamefont {Kato}}]{PhysRevX.4.011005}%
	\BibitemOpen
	\bibfield  {author} {\bibinfo {author} {\bibfnamefont {A.}~\bibnamefont
			{Yokoyama}}, \bibinfo {author} {\bibfnamefont {M.}~\bibnamefont {Yoshida}},
		\bibinfo {author} {\bibfnamefont {A.}~\bibnamefont {Ishii}},\ and\ \bibinfo
		{author} {\bibfnamefont {Y.~K.}\ \bibnamefont {Kato}},\ }\bibfield  {title}
	{\bibinfo {title} {Giant circular dichroism in individual carbon nanotubes
			induced by extrinsic chirality},\ }\href
	{https://doi.org/10.1103/PhysRevX.4.011005} {\bibfield  {journal} {\bibinfo
			{journal} {Phys. Rev. X}\ }\textbf {\bibinfo {volume} {4}},\ \bibinfo {pages}
		{011005} (\bibinfo {year} {2014})}\BibitemShut {NoStop}%
	\bibitem [{\citenamefont {Fedotov}\ \emph {et~al.}(2006)\citenamefont
		{Fedotov}, \citenamefont {Mladyonov}, \citenamefont {Prosvirnin},
		\citenamefont {Rogacheva}, \citenamefont {Chen},\ and\ \citenamefont
		{Zheludev}}]{Fedotov2006}%
	\BibitemOpen
	\bibfield  {author} {\bibinfo {author} {\bibfnamefont {V.~A.}\ \bibnamefont
			{Fedotov}}, \bibinfo {author} {\bibfnamefont {P.~L.}\ \bibnamefont
			{Mladyonov}}, \bibinfo {author} {\bibfnamefont {S.~L.}\ \bibnamefont
			{Prosvirnin}}, \bibinfo {author} {\bibfnamefont {A.~V.}\ \bibnamefont
			{Rogacheva}}, \bibinfo {author} {\bibfnamefont {Y.}~\bibnamefont {Chen}},\
		and\ \bibinfo {author} {\bibfnamefont {N.~I.}\ \bibnamefont {Zheludev}},\
	}\bibfield  {title} {\bibinfo {title} {{Asymmetric propagation of
				electromagnetic waves through a planar chiral structure}},\ }\href
	{https://doi.org/10.1103/PhysRevLett.97.167401} {\bibfield  {journal}
		{\bibinfo  {journal} {Physical Review Letters}\ }\textbf {\bibinfo {volume}
			{97}},\ \bibinfo {pages} {1} (\bibinfo {year} {2006})},\ \Eprint
	{https://arxiv.org/abs/0604234} {arXiv:0604234 [physics]} \BibitemShut
	{NoStop}%
	\bibitem [{\citenamefont {Fedotov}\ \emph {et~al.}(2007)\citenamefont
		{Fedotov}, \citenamefont {Schwanecke}, \citenamefont {Zheludev},
		\citenamefont {Khardikov},\ and\ \citenamefont {Prosvirnin}}]{Fedotov2007}%
	\BibitemOpen
	\bibfield  {author} {\bibinfo {author} {\bibfnamefont {V.~A.}\ \bibnamefont
			{Fedotov}}, \bibinfo {author} {\bibfnamefont {A.~S.}\ \bibnamefont
			{Schwanecke}}, \bibinfo {author} {\bibfnamefont {N.~I.}\ \bibnamefont
			{Zheludev}}, \bibinfo {author} {\bibfnamefont {V.~V.}\ \bibnamefont
			{Khardikov}},\ and\ \bibinfo {author} {\bibfnamefont {S.~L.}\ \bibnamefont
			{Prosvirnin}},\ }\bibfield  {title} {\bibinfo {title} {{Asymmetric
				transmission of light and enantiomerically sensitive plasmon resonance in
				planar chiral nanostructures}},\ }\href {https://doi.org/10.1021/nl0707961}
	{\bibfield  {journal} {\bibinfo  {journal} {Nano Letters}\ }\textbf {\bibinfo
			{volume} {7}},\ \bibinfo {pages} {1996} (\bibinfo {year} {2007})}\BibitemShut
	{NoStop}%
	\bibitem [{\citenamefont {Schwanecke}\ \emph {et~al.}(2008)\citenamefont
		{Schwanecke}, \citenamefont {Fedotov}, \citenamefont {Khardikov},
		\citenamefont {Prosvirnin}, \citenamefont {Chen},\ and\ \citenamefont
		{Zheludev}}]{Schwanecke2008}%
	\BibitemOpen
	\bibfield  {author} {\bibinfo {author} {\bibfnamefont {A.~S.}\ \bibnamefont
			{Schwanecke}}, \bibinfo {author} {\bibfnamefont {V.~A.}\ \bibnamefont
			{Fedotov}}, \bibinfo {author} {\bibfnamefont {V.~V.}\ \bibnamefont
			{Khardikov}}, \bibinfo {author} {\bibfnamefont {S.~L.}\ \bibnamefont
			{Prosvirnin}}, \bibinfo {author} {\bibfnamefont {Y.}~\bibnamefont {Chen}},\
		and\ \bibinfo {author} {\bibfnamefont {N.~I.}\ \bibnamefont {Zheludev}},\
	}\bibfield  {title} {\bibinfo {title} {{Nanostructured metal film with
				asymmetric optical transmission}},\ }\href
	{https://doi.org/10.1021/nl801794d} {\bibfield  {journal} {\bibinfo
			{journal} {Nano Letters}\ }\textbf {\bibinfo {volume} {8}},\ \bibinfo {pages}
		{2940} (\bibinfo {year} {2008})}\BibitemShut {NoStop}%
	\bibitem [{\citenamefont {Aba}\ \emph {et~al.}(2018)\citenamefont {Aba},
		\citenamefont {Qu}, \citenamefont {Wang}, \citenamefont {Chen}, \citenamefont
		{Li}, \citenamefont {Wang}, \citenamefont {Bai},\ and\ \citenamefont
		{Zhang}}]{Aba2018}%
	\BibitemOpen
	\bibfield  {author} {\bibinfo {author} {\bibfnamefont {T.}~\bibnamefont
			{Aba}}, \bibinfo {author} {\bibfnamefont {Y.}~\bibnamefont {Qu}}, \bibinfo
		{author} {\bibfnamefont {T.}~\bibnamefont {Wang}}, \bibinfo {author}
		{\bibfnamefont {Y.}~\bibnamefont {Chen}}, \bibinfo {author} {\bibfnamefont
			{H.}~\bibnamefont {Li}}, \bibinfo {author} {\bibfnamefont {Y.}~\bibnamefont
			{Wang}}, \bibinfo {author} {\bibfnamefont {Y.}~\bibnamefont {Bai}},\ and\
		\bibinfo {author} {\bibfnamefont {Z.}~\bibnamefont {Zhang}},\ }\bibfield
	{title} {\bibinfo {title} {{Tunable asymmetric transmission through tilted
				rectangular nanohole arrays in a square lattice}},\ }\href
	{https://doi.org/10.1364/oe.26.001199} {\bibfield  {journal} {\bibinfo
			{journal} {Optics Express}\ }\textbf {\bibinfo {volume} {26}},\ \bibinfo
		{pages} {1199} (\bibinfo {year} {2018})}\BibitemShut {NoStop}%
	\bibitem [{\citenamefont {Khanikaev}\ \emph {et~al.}(2016)\citenamefont
		{Khanikaev}, \citenamefont {Arju}, \citenamefont {Fan}, \citenamefont
		{Purtseladze}, \citenamefont {Lu}, \citenamefont {Lee}, \citenamefont
		{Sarriugarte}, \citenamefont {Schnell}, \citenamefont {Hillenbrand},
		\citenamefont {Belkin} \emph {et~al.}}]{khanikaev2016experimental}%
	\BibitemOpen
	\bibfield  {author} {\bibinfo {author} {\bibfnamefont {A.~B.}\ \bibnamefont
			{Khanikaev}}, \bibinfo {author} {\bibfnamefont {N.}~\bibnamefont {Arju}},
		\bibinfo {author} {\bibfnamefont {Z.}~\bibnamefont {Fan}}, \bibinfo {author}
		{\bibfnamefont {D.}~\bibnamefont {Purtseladze}}, \bibinfo {author}
		{\bibfnamefont {F.}~\bibnamefont {Lu}}, \bibinfo {author} {\bibfnamefont
			{J.}~\bibnamefont {Lee}}, \bibinfo {author} {\bibfnamefont {P.}~\bibnamefont
			{Sarriugarte}}, \bibinfo {author} {\bibfnamefont {M.}~\bibnamefont
			{Schnell}}, \bibinfo {author} {\bibfnamefont {R.}~\bibnamefont
			{Hillenbrand}}, \bibinfo {author} {\bibfnamefont {M.}~\bibnamefont {Belkin}},
		\emph {et~al.},\ }\bibfield  {title} {\bibinfo {title} {Experimental
			demonstration of the microscopic origin of circular dichroism in
			two-dimensional metamaterials},\ }\href@noop {} {\bibfield  {journal}
		{\bibinfo  {journal} {Nature communications}\ }\textbf {\bibinfo {volume}
			{7}},\ \bibinfo {pages} {1} (\bibinfo {year} {2016})}\BibitemShut {NoStop}%
	\bibitem [{\citenamefont {Krasavin}\ \emph {et~al.}(2005)\citenamefont
		{Krasavin}, \citenamefont {Schwanecke}, \citenamefont {Zheludev},
		\citenamefont {Reichelt}, \citenamefont {Stroucken}, \citenamefont {Koch},\
		and\ \citenamefont {Wright}}]{krasavin2005polarization}%
	\BibitemOpen
	\bibfield  {author} {\bibinfo {author} {\bibfnamefont {A.}~\bibnamefont
			{Krasavin}}, \bibinfo {author} {\bibfnamefont {A.}~\bibnamefont
			{Schwanecke}}, \bibinfo {author} {\bibfnamefont {N.}~\bibnamefont
			{Zheludev}}, \bibinfo {author} {\bibfnamefont {M.}~\bibnamefont {Reichelt}},
		\bibinfo {author} {\bibfnamefont {T.}~\bibnamefont {Stroucken}}, \bibinfo
		{author} {\bibfnamefont {S.~W.}\ \bibnamefont {Koch}},\ and\ \bibinfo
		{author} {\bibfnamefont {E.~M.}\ \bibnamefont {Wright}},\ }\bibfield  {title}
	{\bibinfo {title} {Polarization conversion and focusing of light propagating
			through a small chiral hole in a metallic screen},\ }\href@noop {} {\bibfield
		{journal} {\bibinfo  {journal} {Applied Physics Letters}\ }\textbf {\bibinfo
			{volume} {86}},\ \bibinfo {pages} {201105} (\bibinfo {year}
		{2005})}\BibitemShut {NoStop}%
	\bibitem [{\citenamefont {Krasavin}\ \emph {et~al.}(2006)\citenamefont
		{Krasavin}, \citenamefont {Schwanecke},\ and\ \citenamefont
		{Zheludev}}]{Krasavin_2006}%
	\BibitemOpen
	\bibfield  {author} {\bibinfo {author} {\bibfnamefont {A.~V.}\ \bibnamefont
			{Krasavin}}, \bibinfo {author} {\bibfnamefont {A.~S.}\ \bibnamefont
			{Schwanecke}},\ and\ \bibinfo {author} {\bibfnamefont {N.~I.}\ \bibnamefont
			{Zheludev}},\ }\bibfield  {title} {\bibinfo {title} {Extraordinary properties
			of light transmission through a small chiral hole in a metallic screen},\
	}\href {https://doi.org/10.1088/1464-4258/8/4/s08} {\bibfield  {journal}
		{\bibinfo  {journal} {Journal of Optics A: Pure and Applied Optics}\ }\textbf
		{\bibinfo {volume} {8}},\ \bibinfo {pages} {S98} (\bibinfo {year}
		{2006})}\BibitemShut {NoStop}%
	\bibitem [{\citenamefont {Reichelt}\ \emph {et~al.}(2006)\citenamefont
		{Reichelt}, \citenamefont {Koch}, \citenamefont {Krasavin}, \citenamefont
		{Moloney}, \citenamefont {Schwanecke}, \citenamefont {Stroucken},
		\citenamefont {Wright},\ and\ \citenamefont {Zheludev}}]{Reichelt2006}%
	\BibitemOpen
	\bibfield  {author} {\bibinfo {author} {\bibfnamefont {M.}~\bibnamefont
			{Reichelt}}, \bibinfo {author} {\bibfnamefont {S.}~\bibnamefont {Koch}},
		\bibinfo {author} {\bibfnamefont {A.}~\bibnamefont {Krasavin}}, \bibinfo
		{author} {\bibfnamefont {J.}~\bibnamefont {Moloney}}, \bibinfo {author}
		{\bibfnamefont {A.}~\bibnamefont {Schwanecke}}, \bibinfo {author}
		{\bibfnamefont {T.}~\bibnamefont {Stroucken}}, \bibinfo {author}
		{\bibfnamefont {E.}~\bibnamefont {Wright}},\ and\ \bibinfo {author}
		{\bibfnamefont {N.}~\bibnamefont {Zheludev}},\ }\bibfield  {title} {\bibinfo
		{title} {{Broken enantiomeric symmetry for electromagnetic waves interacting
				with planar chiral nanostructures}},\ }\href
	{https://doi.org/10.1007/s00340-006-2211-4} {\bibfield  {journal} {\bibinfo
			{journal} {Applied Physics B}\ }\textbf {\bibinfo {volume} {84}},\ \bibinfo
		{pages} {97} (\bibinfo {year} {2006})}\BibitemShut {NoStop}%
	\bibitem [{\citenamefont {Prosvirnin}\ and\ \citenamefont
		{Zheludev}(2005)}]{PhysRevE.71.037603}%
	\BibitemOpen
	\bibfield  {author} {\bibinfo {author} {\bibfnamefont {S.~L.}\ \bibnamefont
			{Prosvirnin}}\ and\ \bibinfo {author} {\bibfnamefont {N.~I.}\ \bibnamefont
			{Zheludev}},\ }\bibfield  {title} {\bibinfo {title} {Polarization effects in
			the diffraction of light by a planar chiral structure},\ }\href
	{https://doi.org/10.1103/PhysRevE.71.037603} {\bibfield  {journal} {\bibinfo
			{journal} {Phys. Rev. E}\ }\textbf {\bibinfo {volume} {71}},\ \bibinfo
		{pages} {037603} (\bibinfo {year} {2005})}\BibitemShut {NoStop}%
	\bibitem [{\citenamefont {Meinzer}\ \emph {et~al.}(2013)\citenamefont
		{Meinzer}, \citenamefont {Hendry},\ and\ \citenamefont
		{Barnes}}]{Meinzer2013}%
	\BibitemOpen
	\bibfield  {author} {\bibinfo {author} {\bibfnamefont {N.}~\bibnamefont
			{Meinzer}}, \bibinfo {author} {\bibfnamefont {E.}~\bibnamefont {Hendry}},\
		and\ \bibinfo {author} {\bibfnamefont {W.~L.}\ \bibnamefont {Barnes}},\
	}\bibfield  {title} {\bibinfo {title} {{Probing the chiral nature of
				electromagnetic fields surrounding plasmonic nanostructures}},\ }\href
	{https://doi.org/10.1103/PhysRevB.88.041407} {\bibfield  {journal} {\bibinfo
			{journal} {Physical Review B - Condensed Matter and Materials Physics}\
		}\textbf {\bibinfo {volume} {88}},\ \bibinfo {pages} {1} (\bibinfo {year}
		{2013})}\BibitemShut {NoStop}%
	\bibitem [{\citenamefont {Hopkins}\ \emph {et~al.}()\citenamefont {Hopkins},
		\citenamefont {Poddubny}, \citenamefont {Miroshnichenko},\ and\ \citenamefont
		{Kivshar}}]{Hopkins2015}%
	\BibitemOpen
	\bibfield  {author} {\bibinfo {author} {\bibfnamefont {B.}~\bibnamefont
			{Hopkins}}, \bibinfo {author} {\bibfnamefont {A.~N.}\ \bibnamefont
			{Poddubny}}, \bibinfo {author} {\bibfnamefont {A.~E.}\ \bibnamefont
			{Miroshnichenko}},\ and\ \bibinfo {author} {\bibfnamefont {Y.~S.}\
			\bibnamefont {Kivshar}},\ }\bibfield  {title} {\bibinfo {title} {Circular
			dichroism induced by fano resonances in planar chiral oligomers},\ }\href
	{https://doi.org/10.1002/lpor.201500222} {\bibfield  {journal} {\bibinfo
			{journal} {Laser \& Photonics Reviews}\ }\textbf {\bibinfo {volume} {10}},\
		\bibinfo {pages} {137}}\BibitemShut {NoStop}%
	\bibitem [{\citenamefont {Banzer}\ \emph {et~al.}(2016)\citenamefont {Banzer},
		\citenamefont {Wo{\'z}niak}, \citenamefont {Mick}, \citenamefont {De~Leon},\
		and\ \citenamefont {Boyd}}]{banzer2016chiral}%
	\BibitemOpen
	\bibfield  {author} {\bibinfo {author} {\bibfnamefont {P.}~\bibnamefont
			{Banzer}}, \bibinfo {author} {\bibfnamefont {P.}~\bibnamefont {Wo{\'z}niak}},
		\bibinfo {author} {\bibfnamefont {U.}~\bibnamefont {Mick}}, \bibinfo {author}
		{\bibfnamefont {I.}~\bibnamefont {De~Leon}},\ and\ \bibinfo {author}
		{\bibfnamefont {R.~W.}\ \bibnamefont {Boyd}},\ }\bibfield  {title} {\bibinfo
		{title} {Chiral optical response of planar and symmetric nanotrimers enabled
			by heteromaterial selection},\ }\href@noop {} {\bibfield  {journal} {\bibinfo
			{journal} {Nature communications}\ }\textbf {\bibinfo {volume} {7}},\
		\bibinfo {pages} {1} (\bibinfo {year} {2016})}\BibitemShut {NoStop}%
	\bibitem [{\citenamefont {Wo{\'z}niak}\ \emph {et~al.}(2018)\citenamefont
		{Wo{\'z}niak}, \citenamefont {De~Leon}, \citenamefont {H{\"o}flich},
		\citenamefont {Haverkamp}, \citenamefont {Christiansen}, \citenamefont
		{Leuchs},\ and\ \citenamefont {Banzer}}]{wozniak2018chiroptical}%
	\BibitemOpen
	\bibfield  {author} {\bibinfo {author} {\bibfnamefont {P.}~\bibnamefont
			{Wo{\'z}niak}}, \bibinfo {author} {\bibfnamefont {I.}~\bibnamefont
			{De~Leon}}, \bibinfo {author} {\bibfnamefont {K.}~\bibnamefont
			{H{\"o}flich}}, \bibinfo {author} {\bibfnamefont {C.}~\bibnamefont
			{Haverkamp}}, \bibinfo {author} {\bibfnamefont {S.}~\bibnamefont
			{Christiansen}}, \bibinfo {author} {\bibfnamefont {G.}~\bibnamefont
			{Leuchs}},\ and\ \bibinfo {author} {\bibfnamefont {P.}~\bibnamefont
			{Banzer}},\ }\bibfield  {title} {\bibinfo {title} {Chiroptical response of a
			single plasmonic nanohelix},\ }\href@noop {} {\bibfield  {journal} {\bibinfo
			{journal} {Optics express}\ }\textbf {\bibinfo {volume} {26}},\ \bibinfo
		{pages} {19275} (\bibinfo {year} {2018})}\BibitemShut {NoStop}%
	\bibitem [{\citenamefont {Drezet}\ \emph {et~al.}(2008)\citenamefont {Drezet},
		\citenamefont {Genet}, \citenamefont {Laluet},\ and\ \citenamefont
		{Ebbesen}}]{Drezet:08}%
	\BibitemOpen
	\bibfield  {author} {\bibinfo {author} {\bibfnamefont {A.}~\bibnamefont
			{Drezet}}, \bibinfo {author} {\bibfnamefont {C.}~\bibnamefont {Genet}},
		\bibinfo {author} {\bibfnamefont {J.-Y.}\ \bibnamefont {Laluet}},\ and\
		\bibinfo {author} {\bibfnamefont {T.~W.}\ \bibnamefont {Ebbesen}},\
	}\bibfield  {title} {\bibinfo {title} {Optical chirality without optical
			activity: How surface plasmons give a twist to light},\ }\href
	{https://doi.org/10.1364/OE.16.012559} {\bibfield  {journal} {\bibinfo
			{journal} {Opt. Express}\ }\textbf {\bibinfo {volume} {16}},\ \bibinfo
		{pages} {12559} (\bibinfo {year} {2008})}\BibitemShut {NoStop}%
	\bibitem [{\citenamefont {Sch\"aferling}\ \emph {et~al.}(2012)\citenamefont
		{Sch\"aferling}, \citenamefont {Dregely}, \citenamefont {Hentschel},\ and\
		\citenamefont {Giessen}}]{PhysRevX.2.031010}%
	\BibitemOpen
	\bibfield  {author} {\bibinfo {author} {\bibfnamefont {M.}~\bibnamefont
			{Sch\"aferling}}, \bibinfo {author} {\bibfnamefont {D.}~\bibnamefont
			{Dregely}}, \bibinfo {author} {\bibfnamefont {M.}~\bibnamefont {Hentschel}},\
		and\ \bibinfo {author} {\bibfnamefont {H.}~\bibnamefont {Giessen}},\
	}\bibfield  {title} {\bibinfo {title} {Tailoring enhanced optical chirality:
			Design principles for chiral plasmonic nanostructures},\ }\href
	{https://doi.org/10.1103/PhysRevX.2.031010} {\bibfield  {journal} {\bibinfo
			{journal} {Phys. Rev. X}\ }\textbf {\bibinfo {volume} {2}},\ \bibinfo {pages}
		{031010} (\bibinfo {year} {2012})}\BibitemShut {NoStop}%
	\bibitem [{\citenamefont {Valev}\ \emph {et~al.}(2013)\citenamefont {Valev},
		\citenamefont {Baumberg}, \citenamefont {Sibilia},\ and\ \citenamefont
		{Verbiest}}]{Valev2013}%
	\BibitemOpen
	\bibfield  {author} {\bibinfo {author} {\bibfnamefont {V.~K.}\ \bibnamefont
			{Valev}}, \bibinfo {author} {\bibfnamefont {J.~J.}\ \bibnamefont {Baumberg}},
		\bibinfo {author} {\bibfnamefont {C.}~\bibnamefont {Sibilia}},\ and\ \bibinfo
		{author} {\bibfnamefont {T.}~\bibnamefont {Verbiest}},\ }\bibfield  {title}
	{\bibinfo {title} {{Chirality and chiroptical effects in plasmonic
				nanostructures: Fundamentals, recent progress, and outlook}},\ }\href
	{https://doi.org/10.1002/adma.201205178} {\bibfield  {journal} {\bibinfo
			{journal} {Advanced Materials}\ }\textbf {\bibinfo {volume} {25}},\ \bibinfo
		{pages} {2517} (\bibinfo {year} {2013})}\BibitemShut {NoStop}%
	\bibitem [{\citenamefont {Zu}\ \emph {et~al.}(2016)\citenamefont {Zu},
		\citenamefont {Bao},\ and\ \citenamefont {Fang}}]{zu2016planar}%
	\BibitemOpen
	\bibfield  {author} {\bibinfo {author} {\bibfnamefont {S.}~\bibnamefont
			{Zu}}, \bibinfo {author} {\bibfnamefont {Y.}~\bibnamefont {Bao}},\ and\
		\bibinfo {author} {\bibfnamefont {Z.}~\bibnamefont {Fang}},\ }\bibfield
	{title} {\bibinfo {title} {Planar plasmonic chiral nanostructures},\
	}\href@noop {} {\bibfield  {journal} {\bibinfo  {journal} {Nanoscale}\
		}\textbf {\bibinfo {volume} {8}},\ \bibinfo {pages} {3900} (\bibinfo {year}
		{2016})}\BibitemShut {NoStop}%
	\bibitem [{\citenamefont {Schreiber}\ \emph {et~al.}(2013)\citenamefont
		{Schreiber}, \citenamefont {Luong}, \citenamefont {Fan}, \citenamefont
		{Kuzyk}, \citenamefont {Nickels}, \citenamefont {Zhang}, \citenamefont
		{Smith}, \citenamefont {Yurke}, \citenamefont {Kuang}, \citenamefont
		{Govorov} \emph {et~al.}}]{schreiber2013chiral}%
	\BibitemOpen
	\bibfield  {author} {\bibinfo {author} {\bibfnamefont {R.}~\bibnamefont
			{Schreiber}}, \bibinfo {author} {\bibfnamefont {N.}~\bibnamefont {Luong}},
		\bibinfo {author} {\bibfnamefont {Z.}~\bibnamefont {Fan}}, \bibinfo {author}
		{\bibfnamefont {A.}~\bibnamefont {Kuzyk}}, \bibinfo {author} {\bibfnamefont
			{P.~C.}\ \bibnamefont {Nickels}}, \bibinfo {author} {\bibfnamefont
			{T.}~\bibnamefont {Zhang}}, \bibinfo {author} {\bibfnamefont {D.~M.}\
			\bibnamefont {Smith}}, \bibinfo {author} {\bibfnamefont {B.}~\bibnamefont
			{Yurke}}, \bibinfo {author} {\bibfnamefont {W.}~\bibnamefont {Kuang}},
		\bibinfo {author} {\bibfnamefont {A.~O.}\ \bibnamefont {Govorov}}, \emph
		{et~al.},\ }\bibfield  {title} {\bibinfo {title} {Chiral plasmonic dna
			nanostructures with switchable circular dichroism},\ }\href@noop {}
	{\bibfield  {journal} {\bibinfo  {journal} {Nature communications}\ }\textbf
		{\bibinfo {volume} {4}},\ \bibinfo {pages} {1} (\bibinfo {year}
		{2013})}\BibitemShut {NoStop}%
	\bibitem [{\citenamefont {Kravets}\ \emph {et~al.}(2018)\citenamefont
		{Kravets}, \citenamefont {Kabashin}, \citenamefont {Barnes},\ and\
		\citenamefont {Grigorenko}}]{Kravets2018}%
	\BibitemOpen
	\bibfield  {author} {\bibinfo {author} {\bibfnamefont {V.~G.}\ \bibnamefont
			{Kravets}}, \bibinfo {author} {\bibfnamefont {A.~V.}\ \bibnamefont
			{Kabashin}}, \bibinfo {author} {\bibfnamefont {W.~L.}\ \bibnamefont
			{Barnes}},\ and\ \bibinfo {author} {\bibfnamefont {A.~N.}\ \bibnamefont
			{Grigorenko}},\ }\bibfield  {title} {\bibinfo {title} {{Plasmonic Surface
				Lattice Resonances: A Review of Properties and Applications}},\ }\href
	{https://doi.org/10.1021/acs.chemrev.8b00243} {\bibfield  {journal} {\bibinfo
			{journal} {Chemical Reviews}\ }\textbf {\bibinfo {volume} {118}},\ \bibinfo
		{pages} {5912} (\bibinfo {year} {2018})}\BibitemShut {NoStop}%
	\bibitem [{\citenamefont {Bin-Alam}\ \emph {et~al.}(2021)\citenamefont
		{Bin-Alam}, \citenamefont {Reshef}, \citenamefont {Mamchur}, \citenamefont
		{Alam}, \citenamefont {Carlow}, \citenamefont {Upham}, \citenamefont
		{Sullivan}, \citenamefont {M{\'e}nard}, \citenamefont {Huttunen},
		\citenamefont {Boyd} \emph {et~al.}}]{bin2021ultra}%
	\BibitemOpen
	\bibfield  {author} {\bibinfo {author} {\bibfnamefont {M.~S.}\ \bibnamefont
			{Bin-Alam}}, \bibinfo {author} {\bibfnamefont {O.}~\bibnamefont {Reshef}},
		\bibinfo {author} {\bibfnamefont {Y.}~\bibnamefont {Mamchur}}, \bibinfo
		{author} {\bibfnamefont {M.~Z.}\ \bibnamefont {Alam}}, \bibinfo {author}
		{\bibfnamefont {G.}~\bibnamefont {Carlow}}, \bibinfo {author} {\bibfnamefont
			{J.}~\bibnamefont {Upham}}, \bibinfo {author} {\bibfnamefont {B.~T.}\
			\bibnamefont {Sullivan}}, \bibinfo {author} {\bibfnamefont {J.-M.}\
			\bibnamefont {M{\'e}nard}}, \bibinfo {author} {\bibfnamefont {M.~J.}\
			\bibnamefont {Huttunen}}, \bibinfo {author} {\bibfnamefont {R.~W.}\
			\bibnamefont {Boyd}}, \emph {et~al.},\ }\bibfield  {title} {\bibinfo {title}
		{Ultra-high-q resonances in plasmonic metasurfaces},\ }\href@noop {}
	{\bibfield  {journal} {\bibinfo  {journal} {Nature communications}\ }\textbf
		{\bibinfo {volume} {12}},\ \bibinfo {pages} {1} (\bibinfo {year}
		{2021})}\BibitemShut {NoStop}%
	\bibitem [{\citenamefont {Kravets}\ \emph {et~al.}(2008)\citenamefont
		{Kravets}, \citenamefont {Schedin},\ and\ \citenamefont
		{Grigorenko}}]{Kravets2008}%
	\BibitemOpen
	\bibfield  {author} {\bibinfo {author} {\bibfnamefont {V.~G.}\ \bibnamefont
			{Kravets}}, \bibinfo {author} {\bibfnamefont {F.}~\bibnamefont {Schedin}},\
		and\ \bibinfo {author} {\bibfnamefont {A.~N.}\ \bibnamefont {Grigorenko}},\
	}\bibfield  {title} {\bibinfo {title} {{Extremely narrow plasmon resonances
				based on diffraction coupling of localized plasmons in arrays of metallic
				nanoparticles}},\ }\href {https://doi.org/10.1103/PhysRevLett.101.087403}
	{\bibfield  {journal} {\bibinfo  {journal} {Physical Review Letters}\
		}\textbf {\bibinfo {volume} {101}},\ \bibinfo {pages} {1} (\bibinfo {year}
		{2008})}\BibitemShut {NoStop}%
	\bibitem [{\citenamefont {Khlopin}\ \emph {et~al.}(2017)\citenamefont
		{Khlopin}, \citenamefont {Laux}, \citenamefont {Wardley}, \citenamefont
		{Martin}, \citenamefont {Wurtz}, \citenamefont {Plain}, \citenamefont
		{Bonod}, \citenamefont {Zayats}, \citenamefont {Dickson},\ and\ \citenamefont
		{G{\'e}rard}}]{khlopin2017lattice}%
	\BibitemOpen
	\bibfield  {author} {\bibinfo {author} {\bibfnamefont {D.}~\bibnamefont
			{Khlopin}}, \bibinfo {author} {\bibfnamefont {F.}~\bibnamefont {Laux}},
		\bibinfo {author} {\bibfnamefont {W.~P.}\ \bibnamefont {Wardley}}, \bibinfo
		{author} {\bibfnamefont {J.}~\bibnamefont {Martin}}, \bibinfo {author}
		{\bibfnamefont {G.~A.}\ \bibnamefont {Wurtz}}, \bibinfo {author}
		{\bibfnamefont {J.}~\bibnamefont {Plain}}, \bibinfo {author} {\bibfnamefont
			{N.}~\bibnamefont {Bonod}}, \bibinfo {author} {\bibfnamefont {A.~V.}\
			\bibnamefont {Zayats}}, \bibinfo {author} {\bibfnamefont {W.}~\bibnamefont
			{Dickson}},\ and\ \bibinfo {author} {\bibfnamefont {D.}~\bibnamefont
			{G{\'e}rard}},\ }\bibfield  {title} {\bibinfo {title} {Lattice modes and
			plasmonic linewidth engineering in gold and aluminum nanoparticle arrays},\
	}\href@noop {} {\bibfield  {journal} {\bibinfo  {journal} {JOSA B}\ }\textbf
		{\bibinfo {volume} {34}},\ \bibinfo {pages} {691} (\bibinfo {year}
		{2017})}\BibitemShut {NoStop}%
	\bibitem [{\citenamefont {Johnson}\ and\ \citenamefont
		{Christy}(1972)}]{Johnson1972}%
	\BibitemOpen
	\bibfield  {author} {\bibinfo {author} {\bibfnamefont {P.~B.}\ \bibnamefont
			{Johnson}}\ and\ \bibinfo {author} {\bibfnamefont {R.~W.}\ \bibnamefont
			{Christy}},\ }\bibfield  {title} {\bibinfo {title} {{Optical constants of the
				noble metals}},\ }\bibfield  {journal} {\bibinfo  {journal} {Physical Review
			B}\ }\href {https://doi.org/10.1103/PhysRevB.6.4370}
	{10.1103/PhysRevB.6.4370} (\bibinfo {year} {1972}),\ \Eprint
	{https://arxiv.org/abs/arXiv:1011.1669v3} {arXiv:arXiv:1011.1669v3}
	\BibitemShut {NoStop}%
	\bibitem [{\citenamefont {Schulz}\ \emph {et~al.}(2015)\citenamefont {Schulz},
		\citenamefont {Upham}, \citenamefont {Bouchard}, \citenamefont {De~Leon},
		\citenamefont {Karimi},\ and\ \citenamefont {Boyd}}]{schulz2015quantifying}%
	\BibitemOpen
	\bibfield  {author} {\bibinfo {author} {\bibfnamefont {S.~A.}\ \bibnamefont
			{Schulz}}, \bibinfo {author} {\bibfnamefont {J.}~\bibnamefont {Upham}},
		\bibinfo {author} {\bibfnamefont {F.}~\bibnamefont {Bouchard}}, \bibinfo
		{author} {\bibfnamefont {I.}~\bibnamefont {De~Leon}}, \bibinfo {author}
		{\bibfnamefont {E.}~\bibnamefont {Karimi}},\ and\ \bibinfo {author}
		{\bibfnamefont {R.~W.}\ \bibnamefont {Boyd}},\ }\bibfield  {title} {\bibinfo
		{title} {Quantifying the impact of proximity error correction on plasmonic
			metasurfaces},\ }\href@noop {} {\bibfield  {journal} {\bibinfo  {journal}
			{Optical Materials Express}\ }\textbf {\bibinfo {volume} {5}},\ \bibinfo
		{pages} {2798} (\bibinfo {year} {2015})}\BibitemShut {NoStop}%
	\bibitem [{\citenamefont {Gorodetski}\ \emph {et~al.}(2016)\citenamefont
		{Gorodetski}, \citenamefont {Chervy}, \citenamefont {Wang}, \citenamefont
		{Hutchison}, \citenamefont {Drezet}, \citenamefont {Genet},\ and\
		\citenamefont {Ebbesen}}]{Gorodetski:16}%
	\BibitemOpen
	\bibfield  {author} {\bibinfo {author} {\bibfnamefont {Y.}~\bibnamefont
			{Gorodetski}}, \bibinfo {author} {\bibfnamefont {T.}~\bibnamefont {Chervy}},
		\bibinfo {author} {\bibfnamefont {S.}~\bibnamefont {Wang}}, \bibinfo {author}
		{\bibfnamefont {J.~A.}\ \bibnamefont {Hutchison}}, \bibinfo {author}
		{\bibfnamefont {A.}~\bibnamefont {Drezet}}, \bibinfo {author} {\bibfnamefont
			{C.}~\bibnamefont {Genet}},\ and\ \bibinfo {author} {\bibfnamefont {T.~W.}\
			\bibnamefont {Ebbesen}},\ }\bibfield  {title} {\bibinfo {title} {Tracking
			surface plasmon pulses using ultrafast leakage imaging},\ }\href
	{https://doi.org/10.1364/OPTICA.3.000048} {\bibfield  {journal} {\bibinfo
			{journal} {Optica}\ }\textbf {\bibinfo {volume} {3}},\ \bibinfo {pages} {48}
		(\bibinfo {year} {2016})}\BibitemShut {NoStop}%
	\bibitem [{\citenamefont {Revah}\ \emph {et~al.}(2019)\citenamefont {Revah},
		\citenamefont {Yaroshevsky},\ and\ \citenamefont
		{Gorodetski}}]{revah2019spin}%
	\BibitemOpen
	\bibfield  {author} {\bibinfo {author} {\bibfnamefont {M.}~\bibnamefont
			{Revah}}, \bibinfo {author} {\bibfnamefont {A.}~\bibnamefont {Yaroshevsky}},\
		and\ \bibinfo {author} {\bibfnamefont {Y.}~\bibnamefont {Gorodetski}},\
	}\bibfield  {title} {\bibinfo {title} {Spin-locking metasurface for surface
			plasmon routing},\ }\href@noop {} {\bibfield  {journal} {\bibinfo  {journal}
			{Scientific Reports}\ }\textbf {\bibinfo {volume} {9}},\ \bibinfo {pages} {1}
		(\bibinfo {year} {2019})}\BibitemShut {NoStop}%
	\bibitem [{\citenamefont {Bueno}(2000)}]{bueno2000polarimetry}%
	\BibitemOpen
	\bibfield  {author} {\bibinfo {author} {\bibfnamefont {J.~M.}\ \bibnamefont
			{Bueno}},\ }\bibfield  {title} {\bibinfo {title} {Polarimetry using
			liquid-crystal variable retarders: theory and calibration},\ }\href@noop {}
	{\bibfield  {journal} {\bibinfo  {journal} {Journal of Optics A: Pure and
				Applied Optics}\ }\textbf {\bibinfo {volume} {2}},\ \bibinfo {pages} {216}
		(\bibinfo {year} {2000})}\BibitemShut {NoStop}%
	\bibitem [{\citenamefont {Arteaga}\ \emph {et~al.}(2017)\citenamefont
		{Arteaga}, \citenamefont {Nichols},\ and\ \citenamefont
		{Ant{\'o}}}]{arteaga2017back}%
	\BibitemOpen
	\bibfield  {author} {\bibinfo {author} {\bibfnamefont {O.}~\bibnamefont
			{Arteaga}}, \bibinfo {author} {\bibfnamefont {S.~M.}\ \bibnamefont
			{Nichols}},\ and\ \bibinfo {author} {\bibfnamefont {J.}~\bibnamefont
			{Ant{\'o}}},\ }\bibfield  {title} {\bibinfo {title} {Back-focal plane mueller
			matrix microscopy: Mueller conoscopy and mueller diffractrometry},\
	}\href@noop {} {\bibfield  {journal} {\bibinfo  {journal} {Applied Surface
				Science}\ }\textbf {\bibinfo {volume} {421}},\ \bibinfo {pages} {702}
		(\bibinfo {year} {2017})}\BibitemShut {NoStop}%
	\bibitem [{\citenamefont {Wagner}\ \emph {et~al.}(2012)\citenamefont {Wagner},
		\citenamefont {Heerklotz}, \citenamefont {Kortenbruck},\ and\ \citenamefont
		{Cichos}}]{wagner2012back}%
	\BibitemOpen
	\bibfield  {author} {\bibinfo {author} {\bibfnamefont {R.}~\bibnamefont
			{Wagner}}, \bibinfo {author} {\bibfnamefont {L.}~\bibnamefont {Heerklotz}},
		\bibinfo {author} {\bibfnamefont {N.}~\bibnamefont {Kortenbruck}},\ and\
		\bibinfo {author} {\bibfnamefont {F.}~\bibnamefont {Cichos}},\ }\bibfield
	{title} {\bibinfo {title} {Back focal plane imaging spectroscopy of photonic
			crystals},\ }\href@noop {} {\bibfield  {journal} {\bibinfo  {journal}
			{Applied Physics Letters}\ }\textbf {\bibinfo {volume} {101}},\ \bibinfo
		{pages} {081904} (\bibinfo {year} {2012})}\BibitemShut {NoStop}%
\end{thebibliography}

%

\pagebreak
\widetext
\begin{center}
\textbf{\large Supplemental Materials: Lattice-plasmon induced asymmetric transmission in two-dimensional
chiral arrays}
\end{center}
\setcounter{equation}{0}
\setcounter{figure}{0}
\setcounter{table}{0}
\setcounter{page}{1}
\makeatletter
\renewcommand{\theequation}{S\arabic{equation}}
\renewcommand{\thefigure}{S\arabic{figure}}
\renewcommand{\bibnumfmt}[1]{[S#1]}
\renewcommand{\citenumfont}[1]{S#1}
\section{Simulation Results of Gammadion}

\begin{figure*}
\includegraphics[width=15cm]{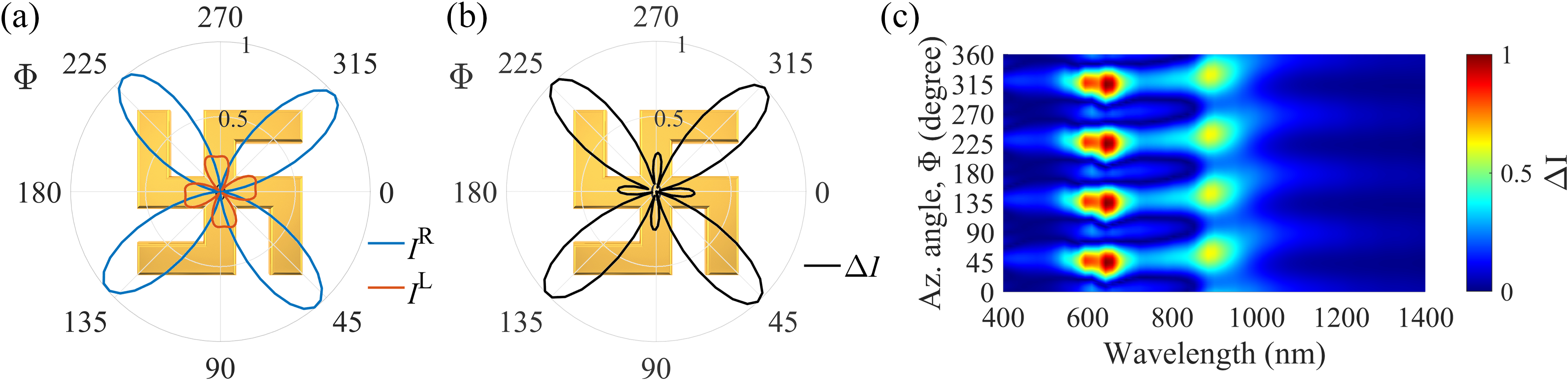}\caption{Optical response of isolated gammadion obtained through FDTD simulations. (a) The polar plot of the farfield scattered intensity as a function of azimuthal angle for a  individual  gammadion nanostructure. $I^{{\rm R}}$, $I^{{\rm L}}$ are the normalized scattered intensity for RCP (solid blue) and LCP (solid red) excitation. (b) The normalized plots of differential in-plane scattering, $\Delta I=\left|I^{{\rm R}}-I^{{\rm L}}\right|,$ for 620 nm excitation wavelength. Superimposed on  the plots are the isolated gammadion nanostructure for  reference. (c) The variation of $\Delta I$ as a function of wavelength for the nanostructure, in normalized units.\label{S:fig:CDS} }
\end{figure*}

Simulations were also carried out for metasurface composed of  squared array of gammadion nanostructure. Although the isolated gammadion has four fold rotation symmetry, it is 2D chiral. Unlike the quadrumer metasurface discussed previously, this makes the gammadion metasurface  2D chiral irrespective of the in-plane rotation of the isolated nanostructure. Fig. \ref{S:fig:CDS}a shows the normalized far field scattered intensity for RCP and LCP excitation at 620 nm. $\Delta I$, which quantifies the difference in angular scattering due to LCP and RCP excitation is shown in Fig. \ref{S:fig:CDS}b. Note that, for this metasurface  in the entire spectral range that $\Delta I_{{\rm max}}$ occurs around odd integer multiples of $45^{\circ}$. As in the case of quadrumer metasurface discussed previously, here too we would see that the angle along which $\Delta I_{{\rm max}}$ occurs play a significant role in the generation of asymmetric transmission. 

\begin{figure*}
\includegraphics[width=15cm]{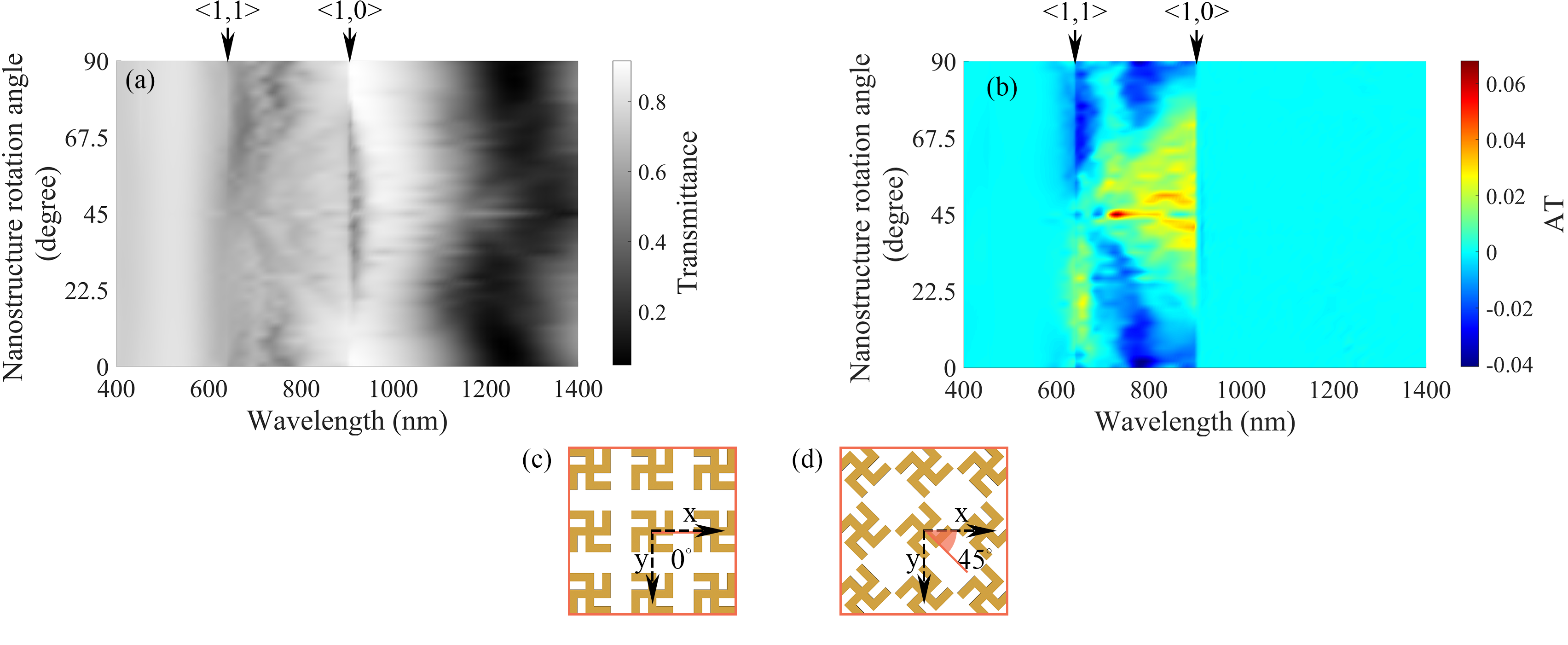}\caption{Optical response of isolated nanostructure obtained through FDTD simulations. (a) The polar plot of the far field scattered intensity as a function of azimuthal angle for an individual  nanostructure. $I^{{\rm R}}$, $I^{{\rm L}}$ are the normalized scattered intensity for RCP (solid blue) and LCP (solid red) excitation. (b) The normalized plots of differential in-plane scattering, $\Delta I=\left|I^{{\rm R}}-I^{{\rm L}}\right|,$ for 620 nm excitation wavelength. Superimposed on  the plots are the isolated nanostructure for  reference. (c) The variation of $\Delta I$ as a function of wavelength for the nanostructure, in normalized units.\label{S:fig:AT} }
\end{figure*}

Figs. \ref{S:fig:AT}a and b show the simulated transmission and asymmetric transmission results for gammadion metasurface as a function of the in-plane rotation of gammadion. The schematic illustration of the metasurfaces obtained by in-plane rotation is given in Figs. \ref{S:fig:AT}c and d. Although the gammadion metasurface shows varying strengths of asymmetric transmission for different in-plane rotation angles of the individual gammadion, the highest asymmetric transmission is obtained when the individual structure is rotated by odd integer multiples of $45^{\circ}$, which coincides with the angles for $\Delta I_{{\rm max}}$ of the isolated gammadion nanostructure (see Fig. \ref{fig:CDS}d). Also, the gammadion rotation angles for which $AT=0$ vary with wavelength, mimicking the characteristics of $\Delta I$ (see in Fig. \ref{fig:CDS}d) of an individual gammadion. From the analysis of both quadrumer and gammadion metasurface it is now clear  that, irrespective of the individual nanostructure being chiral or achiral, the maximum (minimum) asymmetric transmission is observed when the $\Delta I_{{\rm max}}$ ($\Delta I_{{\rm min}}$) is aligned along the corresponding  in-plane diffraction orders of the metasurface.

\section{Fabrication of metasurface}

The quadrumer metasurfaces investigated in this paper were fabricated following a standard metal lift-off procedure. A glass cover slip (170 $\text{\ensuremath{\mu}}$m thick, BK7) is thoroughly cleaned and treated with an oxygen plasma. PMMA is spun onto the sample, the pattern is exposed using e-beam lithography, and is subsequently developed at room temperature. The shapes of the masks were  optimized using shape-correction \citep{schulz2015quantifying}. A thin layer of gold (20 nm) is deposited using thermal evaporation. Note that no additional adhesion layer is used. The lift-off procedure consists of an overnight soak in acetone at $70^{\circ}$C.

\section{Experimental methods for the measurements of asymmetric transmission}
For experimentally observing asymmetric transmission, we use a customized experimental setup operating in a mode similar to leakage radiation microscopy \citep{Gorodetski:16,revah2019spin}. We perform measurements in real and Fourier space to detect the propagating surface modes as well as the diffraction orders. To this end, we perform back focal plane (k-space) polarimetry \citep{bueno2000polarimetry,arteaga2017back,wagner2012back} with a polarimetric analysis. We use a broadband supercontinuum light source (NKT Photonics SuperK Extreme) with wavelength selective filtering accomplished by an acousto-optical tunable filter in the range of 630 nm~--~950 nm (AOTF; Gooch \& Housego AOTFnC-NIR; spectral width $\sim$12 nm) as shown in Fig.~\ref{fig:EXP1}a. 

The spectrally filtered light is fed into a single-mode optical fiber, which acts as a mode filter. The resulting beam with a near fundamental Gaussian profile is then further expanded with the help of two lenses. We use a combination of a broadband polarizer and a quarter-wave plate to control the polarization of the incoming light. The incoming circularly polarized beam is focused using a lens with a focal length of 60 mm, resulting in a focal spot of  $\sim$50 $\mu$m in diameter. The fabricated sample consists of quadrumer arrays with lateral dimensions of 200~$\mu$m by 200~$\mu$m with a periodicity of $\Lambda=600$~nm. The sample is immersed in oil with refractive index matching that of the glass substrate (N-BK7), providing a symmetric refractive index environment for the metasurface. The precise positioning of the sample at the focal plane was done by using a 3D piezo scanning system. The transmitted light from the metasurface was collected by an oil-immersion microscope objective with a numerical aperture (NA) of 1.3. The back focal plane  of this collection lens is then imaged onto a CCD-camera with 12-bit dynamic range to get access to the angular spectrum (Fourier space) of the transmitted light using an additional achromatic lens ($f=300$~mm). A sketch of the experimental setup is shown in Fig. \ref{fig:EXP1}a. In the Fourier space image, transmitted diffraction orders are naturally separated. We record Fourier space images for incident light of both polarization handedness to extract asymmetric transmission based on diffraction orders. To extract meaningful data with respect to the asymmetric transmission from the quadrumer array, the focused beam must be significantly smaller than the lateral extent of the array itself to avoid edge effects from the boundaries of quadrumer array and at the same time sufficiently paraxial to not illuminate the metasurface with a large angular spectrum of plane waves. In the recorded back focal plane images (see Fig. \ref{fig:EXP1}c), the large intensity in the central angular range, corresponds to the zeroth diffraction order mode. We also find wavelength-dependent higher diffraction orders in the back focal plane images. As expected, the NA of the collection microscopic objective sets an upper bound with respect to the collectible diffraction orders. For a sample immersed into index matching oil ($n=1.51$), we define the maximum normalized transverse wave-vector component as $k_{{\rm max}}/k_{0}=1.3/1.51=0.861$, equivalent to a maximum collection angle of microscope objective of $60.25^{\circ}$. It is important to highlight here that we expect to observe certain diffraction orders depending  on the periodicity, the operating wavelength and the microscope objective collection angle. As mentioned earlier, our experimental scheme is similar to leakage radiation microscopy in the Fourier domain, performed for a normally incident weakly focused beam. Due to the angular limitation of the collection microscope objective, the asymmetric transmission effects in higher diffraction orders near the wavelength for the Rayleigh anomaly cannot be observed, as they propagate parallel to the metasurface (in-plane). Hence, we observe the asymmetric transmission at wavelengths around the Rayleigh anomaly qualitatively by performing leakage radiation microscopy in real space in reflection by imaging the metasurface plane. The asymmetric transmission effect can be observed as a difference in intensity of surface waves generated by right- and left-hand circularly polarized light. This was done by placing an imaging lens in the reflection arm and recording the real space image with a CCD camera. 

\end{document}